\documentclass[sort&compress,preprint,12pt]{elsarticle}
\usepackage{amsmath,amsthm,amssymb}
\usepackage{subfig}

\providecommand\bs[1]{\boldsymbol{#1}}

\newcommand{\transp}{\dag}
\newcommand{\BBR}{\mathbb{R}}
\newcommand{\Prob}{\mathrm{Prob}\,}
\newcommand{\oh}{\frac{1}{2}}
\newcommand{\expe}{\mathrm{e}}
\newcommand{\ds}{\displaystyle}
\newcommand{\de}{\ensuremath{\mathrm{d}}}
\newcommand{\dx}{\mathrm{d}x}

\newcommand{\dt}{\mathrm{d}t}
\newcommand{\e}[1]{\mathrm{e}^{#1}}
\newcommand{\ueff}{\mathrm{U}}
\newcommand{\deff}{\mathrm{D}}

\newcommand{\phip}{\phi^{\prime}}
\newcommand{\Vamp}{\bar{\phi}}
\newcommand{\tp}{t^{\prime}}     
 \newcommand{\Ueff}{\mathrm{U}}

 \newcommand{\Deff}{\mathrm{D}}
 \newcommand{\Genf}{\mathcal{L}_f}
\newcommand{\Wf}{W_F}
\newcommand{\Fmc}{F^{\sharp}}
\newcommand{\Smc}{S_F^{\sharp}}
\newcommand{\Xmc}{X^{\sharp}}
\newcommand{\np}{n^{\prime}}
\newcommand{\Transmatnnp}{K_{n\np}}
\newcommand{\Transmatin}{K_{\np n}}
\newcommand{\Transmatnn}{k_n}

\newcommand{\Mwpesup}{\mathcal{M}_W}

\newcommand{\rhoresc}{\tilde{\rho}_{\mathrm{A}}}
\newcommand{\rhorescd}{\tilde{\rho}_{\mathrm{D}}}
\newcommand{\rhoms}{\tilde{\rho}^{(\epsilon)}_{\mathrm{MS,A}}}
\newcommand{\rhomsd}{\tilde{\rho}^{(\epsilon)}_{\mathrm{MS,D}}}
\newcommand{\gpart}{g_{\mathrm{p}}}
\newcommand{\rhomso}{\tilde{\rho}_{A0}}
\newcommand{\rhomsa}{\tilde{\rho}_{A1}}
\newcommand{\rhomsdo}{\tilde{\rho}_{D0}}
\newcommand{\rhomsda}{\tilde{\rho}_{D1}}
\newcommand{\rhomsdb}{\tilde{\rho}_{D2}}
\newcommand{\rhoaamp}{c_1}

\newcommand{\fdic}{\bar{f}}

\newcommand{\Diffref}{\bar{\mathrm{D}}}

\newcommand{\imi}{\mathrm{i}}

\newcommand{\varmatrix}[1]{\ensuremath{\ds \mathit{\mathsf{#1}}}}
\newcommand{\Transmat}{\varmatrix{K}}

\journal{Journal of Computational Physics}

\begin{document}

\begin{frontmatter}

\title{Numerical Methods for Computing Effective Transport Properties of Flashing Brownian Motors\tnoteref{t1}}
\tnotetext[1]{The research of JCL was supported by the DFG Research Center {\sc Matheon} ``Mathematics for Key Technologies'' (FZT86) in Berlin and partially supported by NSF CAREER grant DMS-0449717.   The work of PRK was partially supported by NSF CAREER grant DMS-0449717.   PRK wishes to thank the Zentrum f\"{u}r Interdisziplin\"{a}re Forschung (ZiF) for its hospitality and support during its  ``Stochastic Dynamics:  Mathematical Theory and Applications" program, at which part of this work was completed.  The research of GP is partially supported by the EPSRC, Grant No. EP/H034587 and EP/J009636/1. GP wishes to thank the Biocomputing group at the Institute of Mathematics, FU Berlin for its hospitality and support during a visit at which part of this work was completed.}
\author[fu]{Juan C. Latorre}
\ead{jlatorre@zedat.fu-berlin.de}
\author[rpi,zif]{Peter R. Kramer\corref{cor1}}
\ead{kramep@rpi.edu}
\author[imp]{Grigorios A. Pavliotis}
\ead{g.pavliotis@imperial.ac.uk}
\address[fu]{Institute of Mathematics, Freie Universit\"{a}t Berlin, Arnimallee 6, 14195 Berlin, Germany}
\address[rpi]{Department of Mathematical Sciences, Rensselaer Polytechnic Institute, 110 8th St., Troy, NY 12180}
\address[zif]{Zentrum f\"{u}r Interdisziplin\"{a}re Forschung, Wellenberg 1, 33615 Bielefeld, Germany}
\address[imp]{Department of Mathematics,
South Kensington Campus,
Imperial College London
London SW7 2AZ}

\cortext[cor1]{Corresponding author:  Phone +001 (518) 276-6896, Fax +001 (518) 276-4824}

\begin{abstract}
We develop a numerical algorithm for computing the effective drift and diffusivity of the steady-state behavior of an overdamped particle driven by a periodic potential whose amplitude is modulated in time by multiplicative noise and forced by additive Gaussian noise (the mathematical structure of a flashing Brownian motor). The numerical algorithm is based on a spectral decomposition of the solutions to two equations arising from homogenization theory:  the stationary Fokker-Planck equation with periodic boundary conditions and a cell problem taking the form of a generalized Poisson equation. We also show that the numerical method of Wang, Peskin, Elston (WPE, 2003) for computing said quantities is equivalent to that resulting from homogenization theory. We show how to adapt the WPE numerical method to this problem by means of discretizing the multiplicative noise via a finite-volume method into a discrete-state Markov jump process which preserves many important properties of the original continuous-state process, such as its invariant distribution and detailed balance. Our numerical experiments show the effectiveness of both methods, and that the spectral method can have some efficiency advantages when treating multiplicative random noise, particularly with strong volatility.
\end{abstract}



\end{frontmatter}

\section{Introduction}
\label{sec:intro}
We will develop and discuss numerical approaches to computing the long-time effective dynamics of a particle undergoing overdamped dynamics in a periodic potential, randomly modulated as a function of time, and driven additionally by thermal fluctuations.   In order to reduce the number of parameters under consideration, we 
rescale the spatial variable with respect to the period $L $ of the potential
and the temporal variable with respect to the time which the particle takes to fall from a maximum near to the minimum of the potential under the zero temperature dynamics.   The resulting rescaled equation of motion for the particle  in one dimension can then be written in the form~\citep{kramer2010}:
\begin{equation} \label{gauss_ratchet}
\de X(t) = -\phip(X(t))F(t)\dt + \sqrt{2\theta}\de W(t),
\end{equation}
where $ X(t) $ denotes the particle position as a function of time, $ \theta = k_B T/\Vamp$ is the ratio of the thermal energy $ k_B T $ to the amplitude of potential variations $ \Vamp $, $ \phi (x) $ is the periodic potential structure rescaled to have an order unity scale of variation, $ W (t) $ is a standard Brownian motion with $\langle \de W(t) \rangle=0$ and $\langle \de W(t) \de W(\tp) \rangle = \delta (t-\tp) \de t \de \tp $, and $ F(t) $ describes the random temporal modulations of the potential, which may be assumed without loss of generality to have its amplitude (in mean and/or variance) normalized as desired. 
For simplicity, we will restrict attention to the situation in which $ F(t) $ is an autonomously prescribed stochastic process, i.e., its dynamics are determined independently of $ X(t) $.

Equation~(\ref{gauss_ratchet}) is an example of a Brownian motor~\citep{reimann2002}, a class of stochastic systems which are used to characterize and analyze the mechanisms behind the functioning of biological molecular motors~\citep{oster2002}  as well as to design artificial microscale and nanoscale machines~\citep{ph:bm,ph:abm}.  
An overdamped dynamical description (without inertia) is appropriate because of the small length scales and physical values of the other parameters. The periodic environment reflects the  ordered assembly of an extended microtubule, actin fiber, or artificial substrate.  Thermal fluctuations play an important but not entirely dominant role so that $ \theta $ in practice tends to be somewhat but not much less than $1 $.  Temporal modulations in the potential are induced by some external means, which typically does work on the particle.  In biological settings, the potential in question is the binding potential between the molecular motor and the microtubule or actin fibers, and is modulated by chemical processes such as the binding of ATP or release of phosphate, as well as physical processes such as an unbound ``head" of the motor ``searching" for and landing on a binding site~\citep{fjk:mpkss}.  These physical and chemical processes proceed with effectively random delays because they typically rely on some component fluctuating under thermal effects until it manages to achieve a certain state so that the process goes forward.  Consequently, continuous-time Markov chains (or sometimes renewal processes with more general waiting distributions~\citep{mef:smddk}) are often used to describe the modulations $ F(t) $~\citep{abk:skmdp,cm:mmk,jm:mhohm,hq:mtmmm}, with each state corresponding to a possible geometric conformation of the motor (except for its center of mass position, encoded by $ X(t) $).  In synthetic motors, the modulation $ F(t) $ may often be periodic by design~\citep{ph:bm,ph:abm}.   We will here particularly focus on the special case of a continuous modulation by an Ornstein-Uhlenbeck process.  Such continuous stochastic modulations of the potential could, for example, represent interference of the motion of the molecular motor due to other organelles and structures in the cellular environment.  

The equation~\eqref{gauss_ratchet} is a simplified ``amplitude-modulated flashing ratchet'' model which integrates all these physical features in an essentially minimalist way, and its study has played a key role in the development of the theoretical understanding of molecular motors~\citep{reimann2002}.  That said, the equations describing actual molecular motors, taking into account more detail regarding their  mechanochemical dynamics and relevant spatial degrees of freedom, are generally more complex than the flashing ratchet model (\ref{gauss_ratchet})~\citep{Keller:2013,HancockFricks:2012,abk:mm,Bates:2010}.  We nonetheless choose to use the simple model (\ref{gauss_ratchet}) to most clearly explain the foundational issues regarding a new simulation approach for molecular motors models governed at least in part by continuous-state stochastic processes.  The efforts needed to extend the methodology developed here to more general molecular motor models will be briefly considered in Section~\ref{sec:conc}.

A central practical question in the theory of Brownian motors is the overall long-time behavior of the particle.  The periodicity of the potential and statistical stationarity, or time-periodicity,  of its modulations imply, through a central limit theorem argument~\citep{elston2000}, that the statistics of $ X(t) $ at long time are Gaussian and characterized completely by the mean drift 
\begin{equation}
\Ueff \equiv \lim _{t \rightarrow \infty} \frac{\langle X(t) \rangle}{t} \label{eq:drift}
\end{equation}
and diffusivity
\begin{equation}
\Deff \equiv  \lim_{t \rightarrow \infty} \frac{\langle (X(t) - \Ueff t)^2\rangle}{2 t}, \label{eq:diff}
\end{equation}
where $ \langle \cdot \rangle $ denotes a statistical average over all randomness. 
Of particular interest is how these transport parameters, characterizing the large-scale, long-time behavior of the motor, are related to the microscopic design parameters (such as $ \theta $, the structure of the potential $ \phi $, and parameters characterizing the fluctuation $ F(t) $) in the detailed stochastic differential equation model (\ref{gauss_ratchet}).  Analytical approaches are generally only possible in asymptotic  limits, such as adiabatically slow or rapidly fluctuating modulations $ F(t) $~\citep{reimann2002,rozenbaum:2012}.  Some work has pursued such questions through direct Monte Carlo simulations~\citep{jm:mhohm, dd:birmt} of the stochastic differential equation (\ref{gauss_ratchet}).   This approach is, however, rather expensive because the trajectories must be followed through many spatial periods and typically also several realizations.  Moreover, the nature of the Brownian motor, in particular the relevant parameter regime $ \theta \lesssim 1 $, is such that the particle takes a substantial amount of time to hop from one spatial period to another~\citep{mk:nafpe,evstigneev:2009}.  Accurate computations are further hampered by the slow convergence of a Monte Carlo simulation with respect to computational effort (square root accuracy gains with respect to simulation time and/or number of realizations).  

Deterministic numerical approaches can be alternatively developed based on the equivalence of the stochastic differential equation (\ref{gauss_ratchet}) for trajectories and the Fokker-Planck partial differential equation 
\begin{equation}
\partial_t \rho (x,f,t) =- \partial_x\left(-\phip(x)f \rho (x,f,t)\right) + \theta\partial_{xx} \rho (x,f,t) + \Genf^{\ast} \rho(x,f,t),
\label{eq:fpe}
\end{equation}
for the probability density $\rho (x,f,t) $ of the particle position $x$ at time $ t $.
In Eq.~(\ref{eq:fpe}), $ \Genf $ is the infinitesmal generator operator associated to the Markov process $ F(t) $, and $\Genf^{\ast}$ its adjoint.
\citet{mk:nafpe} developed finite element simulations of this equation with adaptive time stepping to achieve and estimate the long-time behavior for several canonical Brownian motor models, but notes that the periodicity  creates some challenges for the implementation due to its disruption of the banded structure of the matrix formed by projecting the evolution operator in Eq.~\eqref{eq:fpe} onto the finite element basis.  
Another tactic is to derive and numerically solve deterministic equations for the effective drift (\ref{eq:drift}) and diffusivity (\ref{eq:diff}) of the Brownian motor.   \citet{wang2003} and~\citet{hw:mcmse} designed an effective approach, 
which we will refer to as the WPE (Wang-Peskin-Elston) method and summarize in Section~\ref{sec:wpe}, for the case in which the potential modulations $ F(t) $ are governed by finite-state Markov chain dynamics (and which can affect the potential more generally than just through its amplitude).   
We in particular found this algorithm to be very efficient in mapping out the dependence of a two-state flashing ratchet model with respect to various underlying parameters~\citep{kramer2010}.  
 Another means of deriving direct deterministic equations for the effective drift and diffusivity is through homogenization theory~\cite{pavliotis2005a,ab:ssdhf}.  The resulting equations will be summarized in Section~\ref{sec:hom}.  Some relative virtues of this approach is that it can be developed in a continuum framework, without committing to any particular discretization in advance, and follows a classical multiscale analysis.  The equations in~\citep{wang2003,hw:mcmse}, on the other hand, are obtained after a particular numerically suitable spatial discretization which together with the finite-state Markov chain structure of the modulations, induce a grand Markov chain structure to the dynamics.  The derivation of the drift and effective diffusivity are obtained then by manipulations of the associated Kolmogorov equations featuring the transition rate matrices obtained by these discretizations.  

The formulas for the drift and the diffusion coefficient obtained using homogenization theory can be rigorously justified~\cite{PapStrVar77,EthKur86}. A natural question is whether other approaches that have been developed for the calculation of the drift and diffusion coefficients lead to formulas that are equivalent, at least in some appropriate asymptotic limit, to the ones obtained from homogenization theory. As examples we mention the calculation of the diffusion coefficient for a Brownian particle in a tilted periodic potential using the mean first passage time (MFPT) approach~\cite{lifson_jackson62,reimann_al02,bl:odttp}  and the calculation of transport coefficients (not only the diffusion coefficient) using the Green-Kubo theory~\cite{ResibDeLeen77}. The equivalence between the MFPT approach and the Green-Kubo theory with homogenization theory were investigated in~\cite{pavliotis2005a} and~\cite{Pavliotis2010}, respectively. One of the goals of the present paper is to investigate the equivalence between the drift and diffusion coefficient formulas for the WPE algorithm and the (discretized) formulas derived from homogenization theory.
The homogenization formulas can in particular be discretized in the same manner as WPE do at the beginning, with the  result that the same discretized equation for the drift is obtained but different equations result for the effective diffusivity.   
After several numerical experiments verified that the two approaches achieved the same answer, we found that the WPE equations could in fact be derived by a variation of the homogenization argument by simply passing at one point to working with the adjoint of an equation.  A unified framework capable of developing both the WPE and homogenization equations will be presented in Section~\ref{sec:eqwpehom}.

Beyond simply providing another, possibly more transparent, framework for deriving the WPE equations, the homogenization approach affords some flexibility in the numerical discretization.  The spatial discretization pursued by~\cite{wang2003,hw:mcmse} is  carefully designed to maintain the important property of detailed balance, and we do not seek to improve on this aspect.  Rather, we consider how the effective drift and diffusivity for the Brownian motor equation (\ref{gauss_ratchet}) can be effectively computed when the temporal modulations of the potential $ F(t) $ are Markovian and continuous in time.  The prototypical example we shall examine is the Ornstein-Uhlenbeck process, which can be described equivalently as a Gaussian stationary random process with mean zero and correlation function
\begin{subequations}
\label{eq:ou}
\begin{equation}
\langle F(\tp) F(\tp+ t) \rangle = \sigma_F^2 \expe^{-t/\tau}
\end{equation}
or as the solution of the stochastic differential equation
\begin{equation} 
\de F(t)=-\frac{1}{\tau}F(t)\dt+\sqrt{\frac{2\sigma_F^2}{\tau}} \de \Wf(t)
\end {equation}
\end{subequations}
with $ F(0) $ chosen as a mean zero Gaussian random variable with variance $ \sigma_F^2 $.
In the above equations, $\Wf(t)$ is another standard Brownian motion independent of $ W(t) $ in (\ref{gauss_ratchet}), $ \sigma_F^2 = \langle F(t)^2 \rangle $ is the variance of $ F(t) $, and $ \tau $ is the correlation time of $ F(t) $.

We will explore a spectral discretization of the state variable $F $ which is closely related to the continued fraction method applied to many stochastic systems in \citet{risken1996} and to a neural network model in~\citet{acebron:2004}, but not, to our knowledge, to flashing ratchet equations (\ref{gauss_ratchet}).    We compare in Section~\ref{sec:num} the relative efficiency of the WPE and spectrally discretized homogenization equations in computing the effective drift and diffusivity for the flashing ratchet (\ref{gauss_ratchet}) with modulations governed by the Ornstein-Uhlenbeck process (\ref{eq:ou}), using Monte Carlo simulations as a point of reference for accuracy.  
This model has been previously studied in the literature for the rapid decorrelation limit $ \tau \downarrow 0 $ and adiabatic limit $ \tau \rightarrow \infty $ (see \cite{reimann2002} and references therein) but, to our knowledge, the systematic computation of the effective diffusivity $\deff$ is new.    We will in particular study how the transport properties of a flashing ratchet with continuous Ornstein-Uhlenbeck modulations (\ref{eq:ou}) compares with that of two-state flashing ratchets with the same correlation time and variance.

We would like to stress that it is not our intent, in using the WPE method as a point of comparison for the numerical method obtained by a spectral discretization of the homogenization equations, to critique the fundamental ideas of the WPE method.  The WPE method rather serves as a thoughtful and relatively well-developed approach to computing the effective drift and diffusivity of model equations  for molecular motors, such as Eq.~\eqref{gauss_ratchet} (and more generally), when the stochastic modulation  (here $ F(t) $) is represented as a finite-state Markov chain.  The main point of concern for our numerical method is the effective computation of effective drift and diffusivity when the modulation $ F(t) $ is given as a continuous-state stochastic process (such as the Ornstein-Uhlenbeck process~\eqref{eq:ou}).  Thus, the real thrust of our comparison of the two numerical approaches is the relative efficiency of numerical methods based on a spectral discretization or a finite-state Markov chain discretization of the continuous-state stochastic modulation  $ F(t) $.  We return in Section~\ref{sec:conc} to discussing the conclusions of the numerical results in light of other aspects of the WPE method.

We note finally that the methods developed here for the one-dimensional Brownian ratchet equation (\ref{gauss_ratchet}) can be generalized in principle to multiple dimensions, with a somewhat greater computational expense and more complex indexing of tensor products of Hermite polynomials.  Other extensions of relevance to molecular motor modeling will be discussed briefly in Section~\ref{sec:conc}.

\section{The Wang-Peskin-Elston Numerical Algorithm}
\label{sec:wpe}
We begin by describing how the ideas from the Wang-Peskin-Elston (WPE) method~\cite{wang2003,hw:mcmse} can be adapted in order to compute the effective drift and diffusivity of the flashing Brownian ratchet (\ref{gauss_ratchet}) modulated by the continuous Markov process $ F(t) $.
First,  $ F(t) $ is approximated by a finite state, continuous-time Markov chain $ \Fmc (t) $ with state space $ \{f_n\}_{n \in \Smc} $ and transition rate matrix $ \Transmat $ satisfying the property that all the row sums are zero $ \sum_{\np \in \Smc} \Transmatnnp =0 $ and all non-diagonal entries are nonnegative.  In the original formulation of the WPE method, $ F(t) $ is assumed to already be such a discrete-state Markov chain.  The additive inverse of the negative diagonal entries, $ -\Transmatnn $, defines the rate of leaving state $ n $ (inverse of the expected occupancy time), and the non-negative off-diagonal entries $ \Transmatnnp $ define the proclivity of jumping from state $ n $ to state $\np $ in the sense that $ \Transmatnnp/\Transmatnn $ defines the probability of such a jump whenever the Markov chain leaves state $ n $~\citep{gfl:isp}. This step of discretizing the continuous Markov process $ F(t) $ must be performed carefully and in Subsection~\ref{sec:discapp} we will elaborate on how we do this.  
 Of course, if $ F(t) $ is already a finite-state Markov chain, this step is trivial.  Next, the spatial variable $ x $ is discretized so that the Markov process $ (X(t),F(t)) $ defined in Section~\ref{sec:intro}  is approximated by an extended Markov chain  $ (\Xmc (t),\Fmc (t)) $ on a finite state space (the Cartesian product of the discretized state space of $ X $ and $ F $).   A key element to the WPE framework, particularly when employed for the purpose of trajectory simulation (not our focus here) in non-smooth (i.e., sawtooth) potentials, is the definition of the discretized Markov chain dynamics so that it preserves the detailed balance properties of the original equation (\ref{gauss_ratchet}).  To explain this, we begin with the stochastic differential equation (\ref{gauss_ratchet}) with the Markov process $ F(t) $ replaced by a suitable Markov chain approximation $ \Fmc (t) $:
\begin{equation} \label{eq:flashingSDE}
\de X(t)=-\phip(X(t))\Fmc (t) \dt+\sqrt{2\theta}\de W(t).
\end{equation}
We define $ \rho^n (x,t) $ as the probability density for the semidiscretized process $ (X(t),\Fmc (t))$, $n \in \Smc$, so that for any Borel set $ B \in \BBR $ and $ n \in \Smc$,
\begin{equation*}
\Prob \{X(t) \in B, \Fmc (t) = n\} = \int_{B} \rho^n (x,t) \ \de x. 
\end{equation*}
The Fokker-Planck equation describing its evolution is given by 
\begin{equation*}
\ds \frac{\partial \rho^n(x,t)}{\partial t}= \partial_x\left( \phip (x)f^n\rho^n (x,t)+\theta \partial_x\rho^n(x,t)\right)- k_n \rho_n(x,t)+\sum_{\np \neq n} \Transmatin \rho^{\np} (x,t).
\end{equation*}   
Next $X(t)$ is also approximated (in distribution) by a discrete-state, continuous time Markov chain $ \Xmc (t) $ on a regular spatial grid
$ x_{ij} \equiv j + i \Delta x $, $i=1,\ldots,M_x$, $j \in \mathbb{Z}$, where we are assuming a spatial period of $ 1 $ 
and $ \Delta x = 1/M_x $. In this representation of the grid, the parameter $j$ indexes the real line by cells of length one (the normalized period of the potential $\phi$,) while $i$ indexes the grid points with separation distance $1/M_x$ within each cell. The spatially discretized version of $ \rho^n (x,t) $ is indexed in a somewhat unorthodox way which is convenient for the following developments. The function $ p^n_{i} (j,t) $ is defined to be the probability that the discretized joint process $ (\Xmc (t),\Fmc (t)) $  takes values $ (x_{ij},f^n) $ at time $ t $.  Since this function may also be interpreted as the probability that the semidiscretized process $ (X(t),\Fmc (t)) $ takes values in the set $ (x_{ij}-\oh \Delta x,x_{ij}+\oh \Delta x] \times \{n\} $, it can be related to the probability density $ \rho^n (x,t) $, in an approximate sense owing to the discretized approximation of the dynamics, through:
\begin{equation}
p^n_{i} (j,t) \approx \int_{x_{ij}-\oh \Delta x}^{x_{ij}+\oh \Delta x} \rho^n (x,t) \, \de x \label{eq:probdisc}
\end{equation}
 This probability distribution is then represented in vectorial form 
$\mathbf{p}^n(j,t)=(p_1^n(j,t), p_2^n(j,t),\ldots,p_{M_x}^{n}(j,t))$.  The
discretized equations then read
\begin{eqnarray} \label{wpe_aa} 
\frac{\de \mathbf{p}^n(j,t)}{\de t}&=&\varmatrix{L}^n\mathbf{p}^n(j,t)+\varmatrix{L_{+}^n}\mathbf{p}^n(j-1,t)+\varmatrix{L_{-}^n}\mathbf{p}^n(j+1,t) \\
& & \nonumber \qquad \qquad +\sum_{\np \neq n} \Transmatin \mathbf{p}^{\np}(j,t),
\end{eqnarray}
where the matrices appearing in this equation are given by 
\begin{equation}
\begin{array}{lll}
\ds \left[ \varmatrix{L}^n \right]_{i,i} &=-\ds \left(F^n_{i+1/2}+B^n_{i-1/2}+k_n\right) & \textrm{for } i=1,\ldots,M_x, \\
~\\
\ds \left[ \varmatrix{L}^n \right]_{i-1,i} &=\ds B^n_{i-1/2} & \textrm{for } i=1,\ldots , M_x,\\
~\\
\ds \left[ \varmatrix{L}^n \right]_{i+1,i} &= \ds F^n_{i+1/2}, & \textrm{for } i=1,\ldots , M_x,\\
~\\
\ds \left[ \varmatrix{L}^n \right]_{i,i'} &=0 \quad \textrm{ for $|i - i'| \geq 2$} ,\\
~\\
\ds \left[ \varmatrix{L}_{+}^n \right]_{1,M_x} &=\ds F^n_{M_x+1/2},\quad \textrm{zero else},\\
~\\
\ds \left[ \varmatrix{L}_{-}^n \right]_{M_x,1} &=\ds B^n_{1/2},\quad \textrm{zero else}.
\end{array}
\label{eq:WPEmats}
\end{equation}
Intuitively, the matrix $\varmatrix{L}^n$ is a discretization of the Fokker-Planck operator, while the matrices $\varmatrix{L}_{+}^n$ and $\varmatrix{L}_{-}^n$, acting upon the probability vector $ \varmatrix{p}^n $, are a discretization of the probability flux at the boundaries. 
The terms $F^n_{i+1/2}$ and $B^n_{i+1/2}$ represent the transition rates between adjacent cells, and are chosen such that for any frozen choice of $ \Fmc $, the jump process $\Xmc$, with master equation as given by Eq. (\ref{wpe_aa}) without the $\Transmatin $ term and the $ k_n$ terms in $ \varmatrix{L}^n $, satisfies the detailed balance condition  with respect to the Boltzmann distribution $\e{-\phi(x)/\theta}$. 
We refer the reader to~\citet{wang2003} and~\citet{hw:mcmse} for further details of the numerical method.

From the vectors and matrices defined above, we define supervectors and supermatrices with indices 
$ 1,\ldots,\Mwpesup \equiv M_x \times N_F $, where $ N_F = |\Smc| $,  so that the equation (\ref{wpe_aa}) can be expressed in the following abstract form: 
\begin{equation*}
\frac{\de \mathbf{p}(j,t)}{\de t}=\varmatrix{L} \mathbf{p} (j,t)+\varmatrix{L_{+}}\mathbf{p} (j-1,t)+\varmatrix{L_{-}}\mathbf{p}(j+1,t),
\end{equation*}
More precisely, $ \mathbf{p}^n (j,t) = \left[\mathbf{p}\right]_{n M_x + j} (t) $,  where $ \left[\mathbf{v}\right]_i $ is used to denote the $i$th component of a supervector $ \mathbf{v} $ for later convenience.
Other supermatrices and supervectors are indexed similarly, with  $ \varmatrix{L} $ a supermatrix representing the first and fourth terms in Eq.~(\ref{wpe_aa}) (that is, the dynamics acting within a spatial period), while $ \varmatrix{L}_+ $ and $ \varmatrix{L}_- $ are supermatrices representing, respectively, the second and third terms in Eq.~(\ref{wpe_aa}).  These last two supermatrices will have a block diagonal form since they do not couple across different modulation states.

\citet{wang2003} and~\citet{hw:mcmse}~show that the effective drift and diffusivity are obtained as the unique solutions to the following equations:   
\begin{eqnarray}
\ds \ueff &=& \ds \sum_{i=1}^{\Mwpesup} \left[(\varmatrix{L_+}-\varmatrix{L_-})\mathbf{p^s}\right]_i,\vspace{7pt} \label{eq:uWPE}\\ 
\ds \varmatrix{M}\varmatrix{p^s}&=&0, \ds \quad \varmatrix{M}=\varmatrix{L}+\varmatrix{L_-}+\varmatrix{L_+}, \label{eq:psWPE}
\end{eqnarray}
satisfying the normalization condition 
 $\ds \sum_{i=1}^{\Mwpesup} \left[\mathbf{p^s}\right]_i = 1$, where $ \Mwpesup $ is the total number of discrete states. For the effective diffusivity, one must solve,
 \begin{subequations}
\begin{eqnarray}
\ds \deff &= & \ds \frac{1}{2}\sum_{i=1}^{\Mwpesup} \left[(\varmatrix{L_+}+\varmatrix{L_-})\mathbf{p^s}+2(\varmatrix{L_+}-\varmatrix{L_-})\mathbf{r}\right]_i, \vspace{7pt} \label{eq:dWPE} \\
\ds \varmatrix{M}\mathbf{r}&= & \ds \ueff \mathbf{p^s}-(\varmatrix{L_+}-\varmatrix{L_-})\mathbf{p^s}, \label{eq:rWPE}
\end{eqnarray}
\label{eq:deffwpe}
\end{subequations}
with the normalization condition
\begin{equation}
 \sum_{i=1}^{\Mwpesup} \left[\mathbf{r}\right]_i = 0. \label{eq:rnorm}
 \end{equation}

\section{Homogenization equations for effective drift and diffusion}
\label{sec:hom}
An application of the homogenization formalism from~\citet{pavliotis2005a} yields the following system of equations for the effective drift and diffusivity in the continuously modulated flashing ratchet model (\ref{gauss_ratchet}).   We will not repeat the original derivation here since a unified derivation of the WPE and homogenization equations will be provided in Section~\ref{sec:eqwpehom}.  

First we define the infinitesimal generator of the Markov process $ (X(t),F(t)) $, to be understood as operating on functions defined on $ S_X \times S_F $ where $ S_X $ is the spatial domain, here taken as the unit interval with periodic boundary conditions, and $ S_F $ is the state space of the potential modulations $ F(t) $.  Parameterizing the domain $ S_X $ by the variable $ x $ and $ S_F $ by the variable $ f $, the infinitesimal generator reads:
\begin{displaymath}
\mathcal{L}=-\phip (x)f\partial_x + \theta\partial_{xx} + \Genf
\end {displaymath}
where $ \Genf $ is the infinitesimal generator associated with the Markov process $ F(t) $.  For the Ornstein-Uhlenbeck process (\ref{eq:ou}), the associated state space is $ S_F = \mathbb{R} $ and the associated infinitesimal generator is:  
\begin{equation}
\Genf= \frac{1}{\tau}\left(-f\partial_f + \sigma_F^2 \partial_{ff}\right). \label{eq:genf}
\end{equation}
To compute the effective drift and diffusivity, we first solve the
 stationary Fokker-Planck equation 
\begin{equation} \label{rho_c4}
\mathcal{L}^*\rho(x,f)=0, \quad \rho(x+1,f)=\rho(x,f),
\end {equation}     
where
\begin{displaymath}
\mathcal{L}^*=\partial_x\left( \phip (x)f \cdot\right) + \theta\partial_{xx} + \frac{1}{\tau}\left(\partial_f\left(f \cdot\right) + \sigma_F^2 \partial_{ff}\right)
\end {displaymath}
is the Fokker-Planck operator, defined as the adjoint of the infinitesimal generator.
The stationary solution $\rho(x,f)$ is to have periodic boundary conditions in $x \in [0,1]$ and to satisfy the normalization condition:
\begin{equation}
\int_0^1 \int_{-\infty}^\infty \, \rho(x,f) \, \de f \, \dx =1. \label{eq:rhonorm}
\end{equation}
$ \rho (x,f) $ is here actually a \emph{reduced} probability density~\citep[Sec. 2.4]{reimann2002}  associated with the Markov process $ (X(t),F(t)) $ in that only the relative position of $ X(t) $ with respect to the periodic potential is described; the information concerning the nearest integer to $ X(t) $ is contracted out. In other words, $ \rho(x,f) $ is the joint probability density of $ X(t) - \lfloor X(t) \rfloor $ and $ F(t) $, where $ \lfloor \cdot \rfloor $ is the greatest integer function.
 This contraction of the spatial description to a compact domain is necessary for a nontrivial stationary probability distribution to be defined.  

Once $ \rho $ is computed, the effective drift is obtained rather simply as the drift coefficient in (\ref{gauss_ratchet}) averaged over spatial position with respect to the stationary probability distribution:
\begin{equation} \label{u_c4}
\ueff=\langle -\phip (x)f \rangle_{\rho}
\end{equation}
where:
\begin{equation}
\langle g \rangle_{\rho} \equiv
\int_0^1 \int_{-\infty}^\infty \, g(x,f) \rho(x,f) \, \de f \, \dx.
\end{equation}
Next the following cell problem, which has the form of a Poisson equation, must be solved
\begin{equation}\label{cell_c4}
-\mathcal{L}\chi(x,f)=-\phip (x)f-\ueff, \quad \chi(x+1,f)=\chi(x,f).
\end{equation}
$ \chi $ must satisfy also the condition that it grows sufficiently slowly with respect to $ f $ so that:
\begin{displaymath}
\langle |\chi(x,f)|^2 \rangle_{\rho} <\infty.
\end{displaymath}
Moreover, the operator $ \mathcal{L} $ has a one dimensional kernel of constants; the equation~(\ref{cell_c4}) does satisfy the solvability condition, and we impose an extra condition to fix a unique solution:
\begin{equation}
 \langle \chi \rangle_{\rho}  = 0. \label{eq:chizeroadd}
 \end{equation}
From here, the effective diffusivity $\deff$ is computed as~\citep{PavlVogi08}
\begin{equation} \label{eq:deff}
\deff=\theta+\langle \left( -\phip (x)f-\ueff \right) \chi \rangle_{\rho}+2\theta \langle \partial_x\chi \rangle_{\rho}.
\end{equation}
Note that the choice (\ref{eq:chizeroadd}) does not affect the value of this formula for $ \deff $, as it is invariant under the addition of constants to $ \chi $.

\subsection{Comparison with Wang-Peskin-Elston framework}
We observe first of all some direct similarities between the homogenization equations just derived and those characterizing the Wang-Peskin-Elston framework summarized in Section~\ref{sec:wpe}.  First of all, the equation $\varmatrix{M}\varmatrix{p^s}=0$, together with its normalization can be readily understood as a finite volume discretization of the stationary Fokker-Planck equation (\ref{rho_c4}) with normalization (\ref{eq:rhonorm}), and $ \varmatrix{p^s} $ describes exactly the reduced stationary probability distribution of the discretized system $ (\Xmc (t),\Fmc (t) )$. The formula for the effective drift (\ref{eq:uWPE}) is also a simple discretization of equation (\ref{u_c4}).  On the other hand, the formula (\ref{eq:dWPE}) for the effective diffusivity in the WPE approach does not appear to be a discretized approximation of the equation (\ref{eq:deff}) for the effective diffusivity in the homogenization framework.   In particular, the matrix $\varmatrix{M}$ in (\ref{eq:psWPE}) and (\ref{eq:rWPE}) is a discretization of the Fokker-Planck operator, while in order to find $\deff$ from (\ref{eq:deff}) one must solve a problem involving the adjoint of this operator, as in (\ref{cell_c4}).  In Section \ref{sec:eqwpehom}, however, we will present a unified derivation of both the homogenization and WPE equations which explains the consistency of the results obtained with either method (up to numerical errors incurred by choice of discretization).  

For now, we proceed by noting that the homogenization equations have been derived without any prior discretization, so we may solve the equations for the effective drift and diffusivity using any method we please.  We will next describe a spectral approach similar in spirit to the continued-fraction method of \cite{risken1996} developed in other contexts.

\subsection{Spectral decomposition}
\label{sec:spec}
We now develop a numerical algorithm to solve equations (\ref{rho_c4}) and (\ref{cell_c4}) through a spectral decomposition in terms of Hermite polynomials (in $ f $) and Fourier series (in $ x$) of their respective solutions $\rho(x,f)$ and $\chi(x,f)$. This method is analogous to that presented in \cite{risken1996} for computing the effective drift of a particle on a tilted periodic potential, including its extension to computing effective diffusivity~\citep{PavlVogi08}, and which was employed in~\citet{latorre2013} for numerical comparison against theoretical expansions of the transport coefficients with respect to the strength of the tilt and corresponding corrections to the Einstein relation.
We commence by writing equation (\ref{rho_c4}) as:
\begin{displaymath}
\partial_x\left( \phip (x)f \rho(x,f) \right) + \theta\partial_{xx}\rho(x,f) + \frac{1}{\tau}\Genf^*\rho(x,f)=0,
\end{displaymath} 
where $\Genf^*$ is the adjoint operator of $ \Genf $ given in Eq.~(\ref{eq:genf}).  
The invariant probability density of the $F (t)$ dynamics is given by the solution of
\begin{displaymath}
\Genf^* \rho_{F}(f)=0,
\end{displaymath}
which is readily seen to be $\rho_F(f) = (2\pi \sigma_F^2)^{-1/2} \e{-f^2/(2\sigma_F^2)}$.   We write the reduced stationary distribution for the dynamics $ (X(t),F(t)) $ by factoring out this expression
\begin{displaymath}
\rho(x,f)=\rho_{F}(f)\pi(x|f),
\end{displaymath}
where $ \pi (x|f) $ is just the conditional (reduced probability) density of $ X(t) $ given $ F(t) = f$.  
After substituting this ansatz in equation (\ref{rho_c4}) we are left with
\begin{equation} \label{rho2_c4}
\partial_x\left( \phip (x)f\pi(x|f)+\theta \partial_x\pi(x|f)\right)+\frac{1}{\tau}\Genf \pi(x|f) =0.
\end{equation}
A convenient basis which diagonalizes the $ \Genf $ operator (for Ornstein-Uhlenbeck dynamics (\ref{eq:ou})) and is orthonormal with respect to $ \rho_{F} $ is given by the  
Hermite polynomials $\{H_n(f)\}_{n=0}^{\infty} $, which can be defined through
\begin{displaymath}
H_n(f)=\frac{1}{\sqrt{n!}}h_n(f/\sigma_F),
\end{displaymath}
\begin{displaymath}
h_n(f)=(-1)^n \e{f^2/2}\frac{\mathrm{d}^n}{\mathrm{d}f^n}\left(\e{-f^2/2}\right),
\end{displaymath}
and have the properties:
\begin{displaymath}
\mathcal{L}_{F}H_n(f)=-nH_n(f), \quad n=0,1,2,\ldots
\end{displaymath}
\begin{displaymath}
\langle H_n(f) H_m(f)\rangle_{\rho_{f}}=\int_{-\infty}^\infty H_n(f) H_m(f) \rho_{F}(f) \, \de f =\delta_{nm}.
\end{displaymath}
We now expand the solution $\pi(x,y)$ to the equation (\ref{rho2_c4}) with respect to these Hermite polynomials:
\begin{displaymath}
\pi(x|f)=\sum_{n=0}^\infty \, \pi_n(x)H_n(f).
\end{displaymath}
This representation of $\pi(x|f)$ decomposes equation (\ref{rho2_c4}) into an infinite system of coupled ordinary differential equations, 
\begin{subequations} \label{eq:system_pi}
\begin{equation}\label{eq:pi_0}
\sigma_F \partial_x\left(\phip (x)\pi_1(x) \right)+\theta\partial_{xx}\pi_0(x)=0,
\end{equation}
\begin{equation}\label{eq:pi_n}
\mathcal{L}^-_n\pi_{n+1}(x)+\mathcal{L}_n\pi_{n}(x)+\mathcal{L}^+_n\pi_{n-1}(x)=0, \quad n=1,2,\ldots,
\end{equation}
\end{subequations}
with
\begin{equation}
\begin{array}{rl}
\mathcal{L}^-_n\pi_{n+1}(x)=&\ds \sqrt{(n+1)}\sigma_F \partial_x\left(\phip (x)\pi_{n+1}(x) \right),\\
~\\
\mathcal{L}^+_n\pi_{n-1}(x)=&\ds \sqrt{n}\sigma_F \partial_x\left(\phip (x)\pi_{n-1}(x)\right),\\
~\\
\mathcal{L}_n\pi_n(x)=&\ds \left(\theta\partial_{xx}-n\tau^{-1}\right)\pi_n(x).
\end{array}
\label{eq:lopdefs}
\end{equation}
Since $\pi_n(x,f)$ must be periodic functions, an obvious choice for solving the above infinite system of equations is to express $\pi_n(x)$ in terms of Fourier modes,
\begin{equation}
 \pi_n(x)=\sum_{j=-\infty}^\infty \, \pi_n^j  \e{\mathrm{i}\omega_j x}, \quad  \omega_j=2\pi j. \label{eq:pifour}
\end{equation}
This representation leads to an algebraic system which is solved by truncating the series,
\begin{displaymath}
 \pi(x|f) \approx \sum_{n=0}^{N_s} \sum_{j=-M_s}^{M_s} \, \pi_n^j  \e{\mathrm{i}\omega_j x} H_n(f).
\end{displaymath}
We refer the reader to~\ref{app:spectralMethod} for the details of the algorithm and how the system is solved. In the end, $\ueff$ is computed from Equation~\eqref{u_c4} as 
$$
\begin{array}{rl}
\ds \ueff&=\sigma_F \mathrm{Im} \pi_1^1, 
\end{array}
$$
$\deff$ is computed in a similar way~\citep{PavlVogi08}; we expand the solution $ \chi (x,f) $ with respect to the Hermite polynomials in $ f $ and Fourier series in $ x $, and truncate:
\begin{equation*}
\chi (x,f) \approx \sum_{n=0}^{N_s} \sum_{j=-M_s}^{M_s} \, \chi_n^j  \e{\mathrm{i}\omega_j x} H_n(f).
\end{equation*}
After solving for these expansion coefficients through a projection of the governing equations~\eqref{rho_c4} and~\eqref{cell_c4}, we evaluate the effective diffusivity as
\begin{eqnarray} 
\deff&=&\theta+\imi \frac{ \sigma_F }{2} \sum_{n=0}^N \sqrt{n+1} \left[\sum_{j=-M_s}^{M_s} \chi_{n+1}^j \overline{\pi_n^{j+1}} - \chi_{n+1}^j \overline{\pi_n^{j-1}} + \chi_{n}^j \overline{\pi_{n+1}^{j+1}}-\chi_{n}^j \overline{\pi_{n+1}^{j-1}}\right] \notag \\
& &+4\pi \theta \imi \sum_{n=0}^N \sum_{j=-M}^M j \chi_n^j \overline{\pi_n^j}.
\end{eqnarray}
Details of the derivation can be found in~\ref{app:spectralMethod}.

\section{Equivalence between the WPE Numerical Algorithm and Homogenization Theory}
\label{sec:eqwpehom}
We show in this section how the WPE algorithm described in Section~\ref{sec:wpe} and the homogenization theory in Section~\ref{sec:hom} can be obtained through a unified multiscale derivation, in which one simply chooses at one point between working with an equation or its adjoint, and then a specific choice of discretization.

We begin by considering the flashing ratchet model (\ref{gauss_ratchet}) with arbitrary Markov process $ F(t) $ modulating the potential, and do not yet impose any discretization.  The derivation of the homogenized expressions (\ref{u_c4}) and (\ref{eq:deff}) for the effective transport coefficients in~\citet{pavliotis2005a} pursued  a multiple scale analysis of  the backward-Kolmogorov equation,
\begin{displaymath}
\frac{\partial u(x,f,t)}{\partial t}=\mathcal{L} u(x,f,t).
\end{displaymath}
In order to establish the equivalence between the WPE numerical algorithm and the homogenization theory approach, we find it more convenient to instead apply the multiscale technique to the 
 \emph{forward-Kolmogorov} or Fokker-Planck equation associated with (\ref{gauss_ratchet}), namely
\begin{equation} \label{fke_c3}
\partial_t \rho(x,f,t)=\mathcal{L}^*\rho(x,f,t)=\partial_x\left(\phip (x)f\rho(\cdot)\right) +\theta \partial_{xx}\rho(x,f,t) +\mathcal{L}^*_f\rho(x,f,t),
\end{equation}
where $\mathcal{L}^*_f$ is the adjoint of the infinitesimal generator of $F$ with state space $S_f$. 

We will conduct our theoretical development using formal multiple scales analysis, though 
a rigorous justification, together with a proof of the fact that the diffusion coefficient is finite, can be obtained using tools from stochastic analysis, such as the martingale central limit theorem, together with a careful study of the cell problem~\citep{Landim_al}.

\subsection{Equivalence of Drift Formulas}
We begin by seeking a coarse-grained description on advectively-rescaled large space and time scales
\begin{equation}
\rhoresc (x,f,t) = \epsilon^{-1} \rho \left(x/\epsilon,f,t/\epsilon\right) \label{eq:resca}
\end{equation}
and seek a two-space scale solution of the form
\begin{equation}
\rhoresc (x,f,t) = \left.\rhoms (x,\xi,f,t)\right|_{\xi = x/\epsilon} \label{eq:rhoms}
\end{equation}
with small parameter $ 0 < \epsilon \ll 1 $ denoting the separation of scales between the coarse-grained observation scale and the period, $ \xi $ the small-scale space variable.  We do not include a small-scale time variable because the structure of the dynamics is such that the statistical distribution should approach a quasi-steady state on the small-scales~\citep[Sec. 2.4]{reimann2002}, and we are not interested in resolving the transient evolution of the small-scales from the initial data.
We seek solutions $ \rhoms $ which have periodicity $ \rhoms (x,\xi+1,f,t) = \rhoms (x,\xi,f,t) $  in the small-scale variable $ \xi $ corresponding to the periodicity of the potential.
With the chain rule, we find that a solution of 
\begin{equation}
\begin{split}
\partial_t \rhoms &= \epsilon^{-1} \left[\partial_{\xi} \left(\phip (\xi) f \rhoms\right) + \theta  \partial_{\xi \xi} \rhoms 
+ \mathcal{L}^*_f\rhoms \right] \\
& \qquad  + \left[\phip (\xi) f \partial_x \rhoms + 2\theta \partial_{\xi x} \rhoms\right]
+ \epsilon \theta \partial_{xx} \rhoms \label{eq:rhomspde}
\end{split}
\end{equation}
yields, through Eqs.~(\ref{eq:resca}) and (\ref{eq:rhoms}), a solution to Eq.~(\ref{fke_c3}).  Substituting next a perturbation expansion
\begin{equation*}
\rhoms (x,\xi,f,t) = \rhomso (x,\xi,f,t) + \epsilon \rhomsa (x,\xi,f,t) + \ldots
\end{equation*}
into Eq.~(\ref{eq:rhomspde}),
we obtain the asymptotic hierarchy
\begin{subequations}
\begin{eqnarray}
O(\epsilon^{-1}): & 0 &= \mathcal{L}_0^{\ast} \rhomso, \\
O(1): & \partial_t \rhomso &=   \mathcal{L}_0^{\ast} \rhomsa + \phip(\xi) f \partial_x \rhomso + 2 \theta \partial_{\xi x} \rhomsa, \\
& \vdots & \vdots,
\end{eqnarray} 
\label{eq:rhomshier}
\end{subequations}
where we have defined the fundamental operator on small-scale variables: 
\begin{equation*}
\mathcal{L}_0^{\ast} g \equiv \partial_{\xi} \left(\phip (\xi) f g \right) + \theta  \partial_{\xi \xi} g
+ \mathcal{L}^*_f g.
\end{equation*}
In solving these equations, we use the following solvability condition~\cite{pavliotis2008}:

The equation  
\begin{equation}
\mathcal{L}_0^{\ast} g(\xi,f) =h(\xi,f), \label{eq:lgh}
\end{equation}
has a periodic solution $ g(\xi,f) = g(\xi+1,f) $ only when the solvability condition
\begin{equation*}
\int_0^1 \int_{S_f} h(\xi,f) \, \de f \, \de \xi = 0
\end{equation*}
is satisfied.  When this condition holds, Eq.~(\ref{eq:lgh}) has a one-parameter family of solutions 
$ g(\xi,f) = \gpart (\xi,f) + c \pi_0 (\xi,f) $ where $ c $ is an arbitrary real constant, and $ \pi_0 (\xi,f) $ is defined 
as the unique real, periodic solution $ \pi_0 (\xi,f) = \pi_0 (\xi+1,f) $ of the homogenous equation
\begin{subequations}
\begin{equation}
\mathcal{L}_0^{\ast} \pi_0 = 0
\label{eq:fpepi0}
\end{equation}
with the normalization 
\begin{equation}
\int_0^1 \int_{S_f} \pi_0 (\xi,f) \, \de f \, \de \xi = 1. \label{eq:pi0norm}
\end{equation}
\label{eq:pi0}
\end{subequations}
In the case in which $F$ has discrete state space $S_f$, the integral over $S_f$ should be replaced by a sum over states.  This solvability condition is
derived from the fact that the operator $\mathcal{L}_0^{\ast} $  is elliptic, with one-dimensional null space spanned by $ \pi_0 $  and one-dimensional adjoint null space spanned by constants~\cite{pavliotis2008}.

Applying this solvability condition to the $ O(\epsilon^{-1}) $  equation in Eq.~(\ref{eq:rhomshier}) yields the 
result that $ \rhomso (x,\xi,f,t) = \pi_0 (\xi,f) c (x,t) $ for some function of large-scale variables, $ c(x,t)$, to be determined.  Substituting this expression for $ \rhomso $ into the $O(1) $ equation in Eq.~(\ref{eq:rhomshier}), and then imposing the solvability condition  produces the result that 
\begin{equation*}
\frac{\partial c(x,t)}{\partial t} = \nabla \cdot \left( \ueff  c (x,t)\right)
\end{equation*}
which is of course a simple advection equation
with drift velocity
\begin{displaymath}
\ueff=\int_0^1 \int_{S_F} \, -\phip (\xi)f\pi_0(\xi,f) \de f \, \de \xi .
\end{displaymath}

This recovers the homogenization formula (\ref{u_c4}) for the effective drift of the motor. We show now how upon an  appropriate discretization of physical and state space, the WPE formulas \eqref{eq:uWPE} and~\eqref{eq:psWPE} can be recovered.  First of all, the equation~(\ref{eq:psWPE}) is formally just a discretization of the equation~\eqref{rho_c4} for the stationary distribution of the position of the motor.  To make more precise contact with the choice of supermatrices $ \varmatrix{L}$, $ \varmatrix{L_+}$, and $ \varmatrix{L_-} $ used in~\citet{wang2003} and~\citet{hw:mcmse},
we observe that equation \eqref{eq:fpepi0} can be written as,
\begin{equation}
 \theta\partial_\xi\bigg[ \e{-f \phi(\xi)/\theta}\partial_\xi \big(\e{f\phi(\xi)/\theta}\pi_0(\xi,f)\big) \bigg] + \mathcal{L}^*_f \pi_0(\xi,f) = 0,
 \label{eq:fpeexp}
\end{equation}
A direct finite-volume numerical discretization leads to a numerical scheme of the type developed in~\citep{latorre2011}, and a further consistent approximation of the coefficients in the resulting algebraic equations leads to the 
 equation (\ref{eq:psWPE}) with the forward and backward rates in~\eqref{eq:WPEmats} equal to those in~\citet{wang2003,hw:mcmse}.
 To derive the WPE formula~(\ref{eq:uWPE}) for the effective drift, we integrate Eq.~\eqref{eq:fpeexp} over the modulation variable $ f $, noting the integral over the $ \mathcal{L}^*_f $ term vanishes due to integration of a derivative of a function with decaying values as $ |f| \rightarrow \infty $, to obtain the following relation:
\begin{equation}
 \frac{\de}{\de \xi} \int_{S_F} \theta  \e{-f \phi(\xi)/\theta}\partial_\xi \left(\e{f\phi(\xi)/\theta}\pi_0(\xi,f)\right) \de f =0,
 \label{eq:constflux}
\end{equation}
This implies that the integral, which is nothing but the averaged net spatial flux induced by the stationary distribution at a position $ \xi $, is constant on the cell $ [0,1] $, a statement which can alternatively be derived by physical considerations~\citep{reimann2002}.   We next observe that the expression~\eqref{u_c4}  for $\ueff$ can be written as
\begin{displaymath}
 \ueff = \int_0^1 \, \int_{S_F} \theta  \e{f \phi(\xi)/\theta}\partial_\xi \left(\e{-f\phi(\xi)/\theta}\right) \pi_0(\xi,f) \de f \de \xi ,
\end{displaymath}
which, upon integration by parts and the observation that all factors in the integrand are periodic,
leads to
\begin{equation*}
 \ueff =  - \int_0^1 \, \int_{S_F} \theta  \e{-f \phi(\xi)/\theta}\partial_\xi \left(\e{f\phi(\xi)/\theta} \pi_0(\xi,f)\right) \de f \de \xi .
 \end{equation*}
But by Eq.~\eqref{eq:constflux}, the integrand of the $ \xi $ integral in this expression is independent of $ \xi $, and therefore everywhere equal to its value at the period boundary $ \xi = 1 $, so we may equivalently write
\begin{equation}
 \ueff = - \left[\int_{S_F} \theta  \e{-f \phi(\xi)/\theta}\partial_\xi \left(\e{f\phi(\xi)/\theta} \pi_0(\xi,f)\right) \de f \right] \Big|_{\xi=1}.
 \label{eq:ueffflux}
\end{equation}
In continuous variables, this is precisely the flux of probability through the period boundary at $ \xi= 1 $, which again has a clear physical interpretation~\citep{reimann2002}.  The equation~\eqref{eq:uWPE} can be similarly understood, through the general expression of the supermatrices $ \varmatrix{L_+} $ and $ \varmatrix{L_-} $ in terms of forward and backward transition rates, as the discretized form of the net probability flux through the period boundary at $ \xi = 1 $.   More precisely, we approximate the derivative in Eq.~\eqref{eq:ueffflux} through a 
standard centered finite-difference at the points $\xi=1+\Delta x /2$ and $\xi =1-\Delta x /2$, and discretize the state space $S_F $ as described at the beginning of Section~\ref{sec:wpe} to obtain
an expression of the form
\begin{equation}
 \ueff \approx \sum_{n \in \Smc} \left[ A_{n+}^{M_x} (p^s)^n_{M_x}-A_{n-}^{1} (p^s)^n_1\right] \label{eq:uefffd}
 \end{equation}
 for suitable coefficients $ A_{n+}^{M_x} $ and $ A_{n-}^{1}$.  We have used
 the periodicity of the stationary distribution to identify  $\pi_0(1+\Delta x/2,\cdot) = \pi_0(\Delta x/2,\cdot) $, which is then approximated in standard fashion (Eq.~(\ref{eq:probdisc})) in terms of $ \mathbf{p^s} $ (temporarily relaxing supervector indexing for clarity).
 As before, a further consistent approximation of the coefficients $ A_{n+}^{M_x} $ and $ A_{n-}^{1} $ leads to the WPE formula~\eqref{eq:uWPE} for $ \ueff $, with the supermatrix coefficients defined precisely as in~\citet{wang2003,hw:mcmse}.
 
\subsection{Equivalence of Diffusivity Formulas}
We turn next to the derivation of the effective diffusivity $ \deff $ for homogenization theory  (\ref{eq:deff})   
and the WPE equivalent (\ref{eq:deffwpe}), which unlike the drift, appears not to be a simple discretization of the homogenization formula. To this end, we rescale diffusively to large time and space scales, centered about the net drifting motion:
\begin{displaymath}
\rhorescd (x,f,t) = \epsilon^{-1} \rho ((x+\ueff t)/\epsilon,f,t/\epsilon^2) 
\end{displaymath} 
This change of coordinates reveals the leading order long-time dynamics, with the effects of the drift removed by the re-centering of the spatial coordinate.  These dynamics are expected (and confirmed by the following calculation) to be effective diffusion due to functional central limit theorem considerations arising from the periodic spatial environment.
We  seek a solution  of the form
\begin{equation*}
\rhorescd (x,f,t) = \left. \rhomsd (x,\xi,f,t) \right|_{\xi = (x+ \ueff t)/\epsilon}
\end{equation*}
periodic with respect to the small space variable $ \xi $.  We have introduced here a new small scale variable $ \xi $ which undoes the Galilean transformation and saves us the need to otherwise include a fast time scale describing the trivial advection of the stationary small-scale structure.
By the chain rule, we can generate suitable solutions of Eq.~(\ref{fke_c3}) through periodic solutions to the multiscale transformed Fokker-Planck equation:
\begin{equation} \label{fpe_res}
\begin{split}
\partial_t \rhomsd &= \epsilon^{-2} \left[\partial_{\xi} \left((\phip (\xi)  f \rhomsd\right) + \theta  \partial_{\xi \xi} \rhomsd \right] \\
& \qquad + \epsilon^{-1} \left[ (\phip (\xi)f + \ueff)  \partial_x \rhomsd + 2 \theta \partial_{\xi x} \rhomsd\right]
+ \theta \partial_{xx} \rhomsd.
\end{split}
\end{equation}
Upon substituting the perturbation expansion
\begin{equation} \label{rho_per}
\rhomsd (x,\xi,f,t) =\rhomsdo (x,\xi,f,t) + \epsilon \rhomsda (x,\xi,f,t) + \epsilon^2 \rhomsdb (x,\xi,f,t) + \ldots
\end{equation}
into Eq.~(\ref{fpe_res}), we find by equating equal powers of $\epsilon$, 
\begin{eqnarray}
O(\epsilon^{-2}): &0= & \mathcal{L}^{*}_0\rhomsdo \label{per_1}, \vspace{7pt} \\
O(\epsilon^{-1}): &0= & \mathcal{L}^{*}_0\rhomsda + \mathcal{L}^*_1\rhomsda, \label{per_2} \vspace{7pt} \\
O(1): &\ds \partial_t \rhomsdo =&  \mathcal{L}^{*}_0\rhomsdb + \mathcal{L}^*_1\rhomsda+\theta \partial_{xx}\rhomsdo, \label{per_3} 
\end{eqnarray}
where,
\begin{displaymath}
\begin{array}{rl}
\mathcal{L}^{*}_0&=\partial_\xi \left(\phip (\xi)f \, \cdot \right)+\theta\partial_{\xi \xi}+\mathcal{L}^*_f, \vspace{7pt} \\ 
\mathcal{L}^*_1&=\partial_{x}\left(\left( \phip (\xi)f+\ueff\right) \, \cdot \right)+2\theta\partial_{\xi x}.
\end{array}
\end{displaymath}
Using the same solvability condition as above, the $ O(\epsilon^{-2}) $ equation implies that 
\begin{equation}
\rhomsdo (x,\xi,f,t)=c(x,t)\pi_0(\xi,f), \label{eq:rhomsdosol}
\end{equation}
 where $c(\cdot)$ is a function to be determined, and $ \pi_0 $ is defined as in Eq.~(\ref{eq:pi0}).  Equation (\ref{per_2}) reads,
\begin{equation}
-\mathcal{L}^*_0\rhomsda = \mathcal{L}^*_1\rhomsdo 
= \left[\left(\phip(\xi)f+\ueff\right)\pi_0  + 2 \theta \partial_{\xi} \pi_0\right] \partial_x c,
\label{eq:rhomsda}
\end{equation}
which is automatically solvable since the right hand side satisfies 
\begin{displaymath}
\begin{array}{rl}
\ds \int_0^1 \int_{S_f}\, \mathcal{L}^*_1\rho_0 \, \de f\, \de \xi &= \ds \partial_x c \left( \int_{S_f} \, \left[ \int_{0}^{1}\left(\phip(\xi )f+\ueff\right)\pi_0(\xi,f)\, \de \xi +2\theta\pi_0(\xi,f)\Big|_{\xi=0}^1\right] \, \de f\right) \vspace{7pt} \\
&=0,
\end{array}
\end{displaymath}
because of the periodicity of $\pi_0(\xi,f)$ and the definition (\ref{u_c4}) of $\ueff$. As the variables $ x$ and $ t $ enter as parameters in (\ref{eq:rhomsda}), we can treat the function $ \partial_x c (x,t) $ as a multiplicative parameter, and therefore 
express the solution in the form
\begin{equation}
\rhomsda (x,\xi,f,t)= \pi_0 (\xi,f) \rhoaamp (x,t) + \psi(\xi,f)\partial_xc(x,t). \label{eq:rhomsdasol}
\end{equation}
The first term corresponds to the homogenous solution of Eq.~\eqref{eq:rhomsda}, allowing the free multiplicative constant 
to take the form of an arbitrary (for now) function $\rhoaamp (x,t)$ of the variables not involved in the differential operator $ \mathcal{L}^*_0$.   The second term corresponds to the particular solution of Eq.~\eqref{eq:rhomsda}, which is expressed as $ \partial_x c(x,t) $ (acting as an effective constant) multiplied by  the unique solution of the equation
\begin{equation} \label{cell_fp}
-\mathcal{L}^*_0\psi(\xi,f)=\left(\phip (\xi) f+\ueff\right)\pi_0(\xi,f)+2\theta\partial_\xi\pi_0(\xi,f)
\end{equation}
that is periodic, satisfies $ \langle |\psi|^2 \rangle_{\rho} < \infty $,  and with integral chosen to be:
\begin{equation}
\int_0^1 \int_{S_f} \psi(\xi,f) \, \de f \, \de \xi = - \int_0^1 \int_{S_f} \xi \pi_0(\xi,f) \, \de f \, \de \xi. \label{eq:phinorm}
\end{equation}
We note that any constant could have been chosen on the right hand side, without affecting the subsequent derivation of the formula~\eqref{eq:deffalt} for the effective diffusivity, but our particular choice will facilitate connection with the WPE formula~\eqref{eq:deffwpe}.

This is an adjoint equivalent of the \emph{cell problem} (\ref{cell_c4}) that arises from the homogenization analysis of the backward-Kolmogorov equation.  Upon substituting the results (\ref{eq:rhomsdosol}) and (\ref{eq:rhomsdasol}) into Eq.~(\ref{per_3}), we obtain  
\begin{displaymath}
\begin{array}{rl}
\ds \pi_0(\xi,f)\partial_{t}c(x,t)=&\ds \big(\theta\pi_0(\xi,f)+\left(\phip (\xi)f+\ueff\right)\psi(\xi,f)+2\theta\partial_\xi\psi(\xi,f)\big)  \partial^2_{xx}c(x,t) \vspace{7pt}\\
&\ds + \big(\left(\phip (\xi) f + \ueff\right) \pi_0 (\xi,f) + 2 \theta \partial_\xi \pi_0 (\xi,f)\big) \partial_x c_1 (x,t) + \mathcal{L}^*_0\rhomsdb (\cdot).
\end{array}
\end{displaymath}
The solvability condition for this equation then implies that
 $c(x,t)$ satisfies the \emph{diffusion equation}
\begin{displaymath}
\partial_{t}c(x,t)=\deff \, \partial^2_{xx}c(x,t),
\end{displaymath}
where $\deff$ is given by
\begin{eqnarray}
\deff &=& \int_0^1 \int_{S_f}\, \big(\theta \pi_0 (\xi,f) +\left(\phip(\xi)f+\ueff\right)\psi(\xi,f)+2\theta\partial_\xi\psi(\xi,f)\big)\, \de f\, \de \xi \nonumber \\
&=&\theta + \int_0^1 \int_{S_f}\, \left(\phip (\xi)f+\ueff\right)\psi(\xi,f) \, \de f\, \de \xi,
\label{eq:deffalt}
\end{eqnarray}
where we have used the normalization (\ref{eq:pi0norm}) of $ \pi_0 $ and the periodicity of $ \psi $ in the last equality.
This is a somewhat different expression for $\deff$ than was obtained in Eq.~(\ref{eq:deff}) from the same multiscale technique applied to the backward Kolmogorov equation~\citep{pavliotis2005a}.  

Before showing how the expression (\ref{eq:deffalt}) for the diffusivity is related to the WPE algorithm, we prove directly the equivalence with the equation (\ref{eq:deff}) that arises from the original homogenization theory from~\citet{pavliotis2005a}.  The latter are expressed in terms of an auxiliary field $ \chi (\xi,f) $, which satisfies the cell problem
\begin{displaymath}
-\mathcal{L}_0\chi(\xi,f)=\phip (\xi)f\partial_{\xi} \chi - \theta \partial_{\xi \xi}\chi-\mathcal{L}_f\chi=-\phip (z)f-\ueff,
\end{displaymath}
with periodicity in $ \xi $ and integrability $ \langle |\chi|^2 \rangle_{\rho} < \infty $.

From equation (\ref{eq:deffalt}) we have,
\begin{displaymath}
\begin{array}{rl}
\deff&= \ds \theta + \int_0^1 \int_{S_f}\, \left(\phip (\xi)f+\ueff\right)\psi(\xi,f) \, \de f\, \de \xi,  \vspace{7pt} \\
&=\theta + \ds\int_0^1 \int_{S_f}\, \psi(\xi,f)\mathcal{L}_0\chi(\xi,f)\, \de f\, \de \xi,  \vspace{7pt} \\
&=\ds\theta +\int_0^1 \int_{S_f}\,  \chi(\xi,f)\mathcal{L}_0^*\psi(\xi,f) \, \de f\, \de \xi,  \vspace{7pt} \\
\textrm{from (\ref{cell_fp})} &=\ds\theta + \int_0^1 \int_{S_f}\, -\chi(\xi,f)\Big((\phip (\xi)f+\ueff)\pi_0(\xi,f)+2\theta\partial_\xi\pi_0(\xi,f) \Big) \, \de f\, \de \xi,  \vspace{7pt} \\
&=\ds\theta + \int_0^1 \int_{S_f}\, \Big((-\phip (\xi)f-\ueff)\chi (\xi,f) \pi_0 (\xi,f)\Big) -2\theta\chi (\xi,f) \partial_\xi \pi_0 (\xi,f) \, \de f\, \de \xi,  \vspace{7pt} \\
&=\ds\theta + \int_0^1 \int_{S_f}\, \Big((-\phip (\xi)f-\ueff)\chi (\xi,f) \pi_0 (\xi,f) \Big) +2\theta\pi_0 (\xi,f) \partial_\xi \chi (\xi,f) \, \de f\, \de \xi,  \vspace{7pt} \\
&=\theta+\langle(-\phip (\xi)f-\ueff)\chi\rangle_{\pi_0}+2\theta\langle\partial_{\xi}\chi\rangle_{\pi_0}.
\end{array}
\end{displaymath}
The above expression is precisely equation (\ref{eq:deff}) for $\deff$ found via the original homogenization theory in~\citep{pavliotis2005a}.

The equivalence of the expression (\ref{eq:deffalt}) for the effective diffusivity with that of the WPE numerical method can be established as follows. From equation (\ref{cell_fp}),
\begin{displaymath}
-\mathcal{L}^*_0\psi(\xi,f)=\left(\phip (\xi) f+\ueff\right)\pi_0(\xi,f)+2\theta\partial_\xi\pi_0(\xi,f).
\end{displaymath}
Now define
\begin{equation}
R(\xi,f)=-\left(\psi(\xi,f)+\xi\pi_0(\xi,f)\right). \label{eq:rphipi}
\end{equation}
It is easy to verify that $R$ satisfies the boundary condition 
\begin{equation}
R(\xi+1,f)=R(\xi,f)-\pi_0(\xi,f) \label{eq:rbound}
\end{equation}
 and the equation,
\begin{equation} \label{cont_wpe}
\mathcal{L}^*_0R(\xi,f)=\ueff \pi_0.
\end{equation}
Moreover from Eqs.~\eqref{eq:phinorm} and~\eqref{eq:rhonorm}, we have that
\begin{equation}
 \int_0^1 \, \int_{S_F} R(\xi,f) \de f \de \xi = 0. \label{eq:rzero}
\end{equation}
The same argument based on finite-volume discretizations used to connect the continuous equation \eqref{eq:fpepi0}
with the discretized WPE equation~\eqref{eq:uWPE}, show that~\eqref{eq:rWPE} is simply a discretization of Eq.~\eqref{cont_wpe}.
The matrix $\varmatrix{M}$ is a discretization of the operator $\mathcal{L}_0^*$ corresponding to periodic boundary conditions, and $\varmatrix{r}=\ds \left(R_1,R_2,\ldots,R_{\Mwpesup}\right)^T$, $\varmatrix{p^s}=\ds \left(\pi_1,\pi_2,\ldots,\pi_{\Mwpesup}\right)^T$ are the corresponding discretized approximations of $R(\xi,f)$ and $\pi_0(\xi,f)$. The term  $-(\varmatrix{L_+}-\varmatrix{L_-})\varmatrix{p^s}$ is a correction term arising from the fact that $ R(\xi,f) $ does not satisfy periodic boundary conditions, but rather Eq.~\eqref{eq:rbound}.  Thus, the discretized terms corresponding to ``fluxes of $ R$'' from outside the period domain $ [0,1] $ must involve this shift of $ \pm \pi_0 $ relative to the case in which $ R $ is  periodic and fluxes in from the left/right are equated to fluxes out of the right/left boundary.  Also, the normalization condition~\eqref{eq:rnorm} is clearly a direct discretization of the integral condition~\eqref{eq:rzero}.

Finally, we show how \eqref{eq:deffalt} leads to the WPE formula~\eqref{eq:dWPE} for the effective diffusivity.
First, we note that by periodicity of $ \psi (\xi,f) $ and the integration over a complete spatial period in $ \psi $, as well as Eq.~\eqref{eq:rphipi} and the normalizations~\eqref{eq:rhonorm} and~\eqref{eq:rnorm}, we can rewrite Eq.~\eqref{eq:deffalt} as:
\begin{align}
\deff &= \theta + \int_{0}^1 \int_{S_F} (\phip (\xi) f + \theta \partial_\xi + \ueff) \psi (\xi,f) \de f \de \xi \nonumber \\
&= \theta - \int_{0}^1 \int_{S_F} (\phip (\xi) f + \theta \partial_\xi + \ueff) \left(R(\xi,f) + \xi \pi_0 (\xi,f)\right)\de f \de \xi 
\nonumber \\
&= \theta -   \int_{0}^1 \int_{S_F} (\phip (\xi) f + \theta \partial_\xi) R(\xi,f) \de f \de \xi
- \int_0^1 \int_{S_F} \xi (\phip (\xi) f + \theta \partial_\xi) \pi_0 (\xi,f) \de f \de \xi \nonumber \\
& \qquad \qquad - \int_0^1 \int_{S_F} \theta \pi_0 (\xi,f) \de f \de \xi - \ueff \left[\int_{0}^1 \int_{S_F} R(\xi,f) \de f \de \xi + \int_{0}^1 \int_{S_F} \xi \pi_0 (\xi,f) \de f \de \xi\right] \nonumber \\
&= -  \int_0^1 \int_{S_F}  \theta \left[ \e{-f \phi/\theta} \partial_{\xi} \left( \e{f \phi/\theta} R(\xi,f)\right)\right] \de f \de \xi 
\nonumber \\
& \qquad \qquad-  \int_0^1 \xi \int_{S_F}  \theta  \left[ \e{-f\phi/\theta} \partial_{\xi} \left( \e{f \phi/\theta} \pi_0 (\xi,f)\right)\right] \de f \de \xi 
\nonumber \\
& \qquad \qquad - \ueff \int_{0}^1 \int_{S_F} \xi \pi_0 (\xi,f) \de f \de \xi. 
\label{eq:deffhack}
\end{align}
Next we integrate by parts in the first integral, noting the periodicity of all factors in the integrand except $ R $, which satisfies Eq.~\eqref{eq:rbound}, to obtain:
\begin{eqnarray}
  -\int_0^1 \, \int_{S_F} \theta \left[\e{-f \phi/\theta} \partial_\xi \left( \e{f \phi/\theta} R\right)\right] \de f \de \xi 
 &=& \left[-\xi \int_{S_F} \theta \e{-f \phi/\theta} \partial_\xi \left( \e{f \phi/\theta} R\right) \de f\right]\Big|_0^1 \label{eq:deffhack1} \\
&& +\int_0^1 \, \int_{S_F} \xi \partial_\xi \left(\theta \e{-f \phi/\theta} \partial_\xi \left( \e{f \phi/\theta} R\right)\right) \de f \de \xi \notag \\
\textrm{from \eqref{cont_wpe}}&=&\left[-\int_{S_F} \theta \e{-f \phi/\theta} \partial_\xi \left( \e{f \phi/\theta} R\right) \de f\right]\Big|_{\xi=1} \notag \\
& & +\int_0^1 \, \int_{S_F} \xi \ueff \pi_0 \de f \de \xi. \notag
\end{eqnarray}
Noting from our argument from Eq.~\eqref{eq:constflux}, the integral over $ S_F $ in the second term in the last expression of Eq.~\eqref{eq:deffhack} is in fact constant with respect to $ \xi $, we can trivially integrate over $ \xi $ to obtain:
\begin{equation}
-  \int_0^1 \xi \int_{S_F}  \theta  \left[ \e{-f \phi/\theta} \partial_{\xi} \left( \e{f \phi/\theta} \pi_0 (\xi,f)\right)\right] \de f \de \xi
= -\frac{1}{2} \left[\int_{S_F} \theta \e{-f \phi/\theta} \partial_\xi \left( \e{f \phi/\theta} \pi_0\right) \de f \right]\Big|_{\xi=1}.
\label{eq:deffhack2}
\end{equation}
Combining then Eqs.~\eqref{eq:deffhack}, \eqref{eq:deffhack1}, and~\eqref{eq:deffhack2},  we obtain the following expression for the effective diffusivity in terms of continuum variables which, analogously to the formula Eq.~\eqref{eq:ueffflux} for the effective drift, only involves evaluations of ``fluxes'' at the period boundary rather than integration with respect to the spatial variable $ \xi $:
\begin{equation}
\deff = -\left\{\int_{S_F} \theta \e{-f \phi/\theta} \partial_\xi \left[ \e{f \phi/\theta} \left(R (\xi,f) + \frac{1}{2} \pi_0\right) \right] \de f\right\}\Big|_{\xi=1}. \label{eq:deffflux}
\end{equation}
Were $ \left(R + \frac{1}{2} \pi_0\right)$ a periodic function, then by the same argument as above which interpreted the expression in Eq.~\eqref{eq:ueffflux} as a net spatial flux of probability and the matrices $ \varmatrix{L_+} - \varmatrix{L_-} $ as a corresponding discretization of operators mapping probability densities to rightward and leftward spatial fluxes, we would say that the right hand side of Eq.~\eqref{eq:deffflux} could be discretized as $ \sum_{i=1}^{\Mwpesup} \left[(\varmatrix{L_+} - \varmatrix{L_-}) (\mathbf{r} + \frac{1}{2} \mathbf{p^s})\right]_i$.  But $ R $ is not periodic, and this argument is flawed because while $ \varmatrix{L_+} \mathbf{r} $ does serve as an appropriate discretization of the ``rightward spatial flux'' of $ R $ across $ \xi=1 $,
$ \varmatrix{L_-} \mathbf{r} $ describes the ``leftward spatial flux'' of $ R $ across $ \xi =0 $, which is not the same as the ``leftward spatial flux'' of $ R $ across $ \xi =1 $ due to the lack of periodicity of $ R $.  Rather, since by Eq.~\eqref{eq:rbound}, $ R(1+\xi,f) = R(\xi,f) - \pi_0 (\xi,f) $, we should discretize the ``leftward spatial flux'' of $ R $ across $ \xi=1 $ as $ \varmatrix{L_-} (\mathbf{r} - \mathbf{p^s}) $.  Then taking the net spatial flux at $ \xi = 1 $ as the ``rightward spatial flux'' minus the ``leftward spatial flux,''' integrated over the modulation variable $ f $,  we would discretize Eq.~\eqref{eq:deffflux} as:
\begin{align*}
\deff \approx &\sum_{i=1}^{\Mwpesup} 
\left[\varmatrix{L_+} \mathbf{r} - \varmatrix{L_-} (\mathbf{r}-\mathbf{p^s}) + (\varmatrix{L_+} - \varmatrix{L_-}) \frac{1}{2} \mathbf{p^s}\right]_i \\
& = \frac{1}{2} \sum_{i=1}^{\Mwpesup} \left[2(\varmatrix{L_+} - \varmatrix{L_-}) \mathbf{r} + (\varmatrix{L_+} + \varmatrix{L_-}) \mathbf{p^s}\right]_i,
\end{align*}
in agreement with the WPE expression~\eqref{eq:dWPE}.

We can make this argument somewhat more concrete, as we did for $ \ueff $ around the discussion of Eq.~\eqref{eq:uefffd}, by approximating the last expression in Eq.~\eqref{eq:deffflux} by a centered finite-difference using the points $\xi=1+\Delta x/2$ and $\xi=1-\Delta x /2$, but now we must notice that $R(1+\Delta x/2)=R(\Delta x /2)-\pi_0(\Delta x /2)$ (while $ \pi_0 (1 + \Delta x/2) = \pi_0 (\Delta x/2)$), to obtain:
\begin{eqnarray*}
\deff &\approx& - \sum_{n \in \Smc} \left[A_{n-}^{1}(r^n_1 - (p^s)^n_1 + \frac{1}{2} (p^s)^n_1) - A_{n+}^{M_x} 
(r^n_{M_x} + \frac{1}{2} (p^s)^n_{M_x})\right] \\
&=& \frac{1}{2} \sum_{n \in \Smc} 2 \left(A_{n+}^{M_x} r^n_{M_x} - A_{n-}^{1} r^n_1\right)
+ \left(A_{n+}^{M_x} (p^s)^n_{M_x} + A_{n-}^{1} (p^s)^n_1\right),
\end{eqnarray*}
where we temporarily suspend supervector indexing of  $ \mathbf{r} $ and $ \mathbf{p^s} $.  The same consistent approximation of the coefficients $ A_{n+}^{M_x} $ and $ A_{n-}^{1} $ as in our discussion of the effective drift gives precisely the WPE formula~\eqref{eq:dWPE} for the effective diffusivity, with supermatrix coefficients defined precisely as in~\citet{wang2003,hw:mcmse}.

\section{Numerical Results}
\label{sec:num}
In this section we will explore the efficacy of the WPE and homogenization approaches in simulating a flashing ratchet (\ref{eq:flashingSDE}) where the potential is modulated by an Ornstein-Uhlenbeck process (\ref{eq:ou}).  Though we have shown that the formulas for the effective drift and diffusivity are formally equivalent for the two methods, their implementations differ in their discretization.  In particular, the homogenization algorithm will discretize both space and the random potential modulation through a spectral expansion, as discussed in Section \ref{sec:hom}.  On the other hand, the WPE algorithm will discretize both space and the random potential modulation through regularly spaced grids, in such a way that the stochastic differential system (\ref{eq:flashingSDE}) is approximated by a finite-state Markov chain which preserves detailed balance  (Subsection~\ref{sec:discapp}).  We also examine 
the theoretical question of how the transport properties of the motor compare under continuous-state or discrete-state potential modulations with equivalent low order statistics (Subsection~\ref{sec:disccontcomp}).  The numerical studies for both the comparison of the algorithms and the discrete-state and continuous-state modulations are presented together in Subsection~\ref{sec:wpevsspec}.

\subsection{Discrete-state approximation of the Ornstein-Uhlenbeck process.}
\label{sec:discapp}
The equation of motion of the flashing ratchet is given in Eq.~\eqref{gauss_ratchet}
where $F(t)$ is the external modulation, which we now fix as the Ornstein-Uhlenbeck (OU) process~\eqref{eq:ou}.
The means for computing the effective drift and diffusivity for this system using homogenization theory, including their discrete, computable approximation, were presented in Section \ref{sec:hom}.   
The WPE numerical method for computing the transport coefficients, on the other hand, is formulated in~\citet{wang2003,hw:mcmse}  for flashing ratchets where $F(t)$ is a continuous-time Markov chain with a finite state space. Consequently, to implement this approach on the continuously modulated model (\ref{eq:ou}), we must somehow approximate the continuous dynamics of the OU-process by a finite-state, continuous-time Markov chain. 
Naive discretizations of either the backward-Kolmogorov or forward-Kolmogorov equation with some finite-difference method can give rise to some inconsistencies. For instance, if the grid size is not small enough the jump rates may not be positive, and some important properties of the original continuous process, such as those concerning its invariant distribution, may be lost.  A systematic general framework for suitable consistent numerical approximations of continuous-state stochastic processes by Markov jump processes is presented in~\citet{KushnerDupuis:2012Ch4}.  
Here we will adopt the particular formal finite-volume discretization procedure of~\citet{latorre2011}, which can also be used to derive the WPE scheme and more generally to discretize a certain class of $N$-dimensional stochastic differential equations with respect to non-rectangular cells.  
 We begin by writing the backward-Kolmogorov equation for $F(t)$,
\begin{displaymath}
 \frac{\partial u(f,t)}{\partial t}=\frac{1}{\tau}\left(-f\frac{\partial u}{\partial f}+\sigma_F^2 \frac{\partial^2 u}{\partial f^2}\right),
\end{displaymath}
Defining $\beta=\sigma_F^{-2}$ and $V(f)=f^2/2$, the backward-Kolmogorov equation can be rewritten as
\begin{equation} \label{eq:bke_1}
\frac{\partial u(f,t)}{\partial t}= \ds \frac{\sigma_F^2}{\tau}\e{\beta V(f)}\frac{\partial}{\partial_f} \left( \e{-\beta V(f)} \frac{\partial}{\partial_f} u(f,t) \right).
\end{equation}
Once written in this form, a finite-volume method can be used to discretize the equation. In one dimension and for a uniform grid, a simple finite-difference scheme can be used to obtain the same approximation. This is presented in detail in~\ref{app:ouApproximation}. In the end, the approximation of the backward-Kolmogorov equation can be expressed as,
\begin{displaymath}
\frac{\de}{\dt} \varmatrix{u}(t)=\varmatrix{L} \varmatrix{u}(t),
\end{displaymath}
where the entries of the matrix $\varmatrix{L}$ are given as,
\begin{displaymath}
 [\varmatrix{L}]_{n,\np}=\left\{ \begin{array}{cl}
			  -(K_{n,n+1}+K_{n,n-1}) & \textrm{if } n=\np, \\
                          K_{n,n+1} & \textrm{if } \np=n+1, \\
                          K_{n,n-1} & \textrm{if } \np=n-1, \\
			  0 & \textrm{otherwise.}
                         \end{array} \right.
\end{displaymath}
where the $ K_{n,\np} $ are nonnegative constants with expressions given in Eq.~(\ref{eq:transRatesF}), 
and $\varmatrix{u}(t) = \left( u_1(t), u_2(t), \ldots, u_{N_F}(t)\right)$ is the pointwise approximation of the solution $u(f,t)$. 
The spatial discretization for the WPE method then follows the standard procedure described in~\citet{wang2003,hw:mcmse}, building upon this finite-state Markov chain approximation for $ F(t)$, which we denote $ \Fmc (t) $.   We will refer to the resulting numerical method, extending the WPE ideas with the finite-volume discretization procedure of~\citet{latorre2011} to handle the discretization of the stochastic process $ F(t) $, as the ``WPE-based'' method in the following discussion.

 \subsection{Comparison between discrete-state and continuous-state  flashing ratchet.}
 \label{sec:disccontcomp}
We next turn to the question of how sensitively the transport properties depend on a discretization of  the continuous-state modulation of the flashing ratchet (\ref{eq:ou}) that preserves exactly the most basic low-order statistics, namely the mean and correlation function. 
  Of course we expect that with sufficiently many discrete states, the transport properties should be relatively insensitive to the discretization, so we set the comparison most starkly by comparing the Ornstein-Uhlenbeck modulation with a dichotomous Markov-chain modulation $F_D (t) $ taking values $ \{f_1,f_2\}$ with transition rates between the two states given by $k_{12}$ and $k_{21}$.  We choose these parameters to mimic the Ornstein-Uhlenbeck process as closely as possible.  First, because the OU-process is symmetric about the origin, we set $f_1=-f_2=\fdic$ and $k_{12}=k_{21}=k$.  This makes the dichotomous process have mean zero, as does the OU-process.  We next demand that both the discrete and continuous process have the same correlation function, assuming both are initialized with respect to their stationary distributions.  The OU-process has correlation function~\citep{gardiner2003}:
\begin{equation*}
\langle F (\tp) F(\tp + t) \rangle = \sigma_F^2 \expe^{-t/\tau}
\end{equation*}
whereas the dichotomous Markov chain has correlation function
\begin{equation*}
\langle F_D (\tp) F_D (\tp+t) \rangle = \fdic{}^2 \expe^{-2kt}.
\end{equation*}
 We set then,
\begin{displaymath}
\fdic = \sigma_F, \qquad \qquad  k=\frac{1}{2\tau}.
\end{displaymath}
These restrictions completely determine the Markov chain $F_D$.

\subsection{Comparison of WPE-based and Homogenization Algorithms for Continuous Potential Modulations}
\label{sec:wpevsspec}
\begin{figure}[th]
 \centering
 \begin{subfloat}[$\tau=0.01$]{
  \centering
  \includegraphics[width=0.45\textwidth]{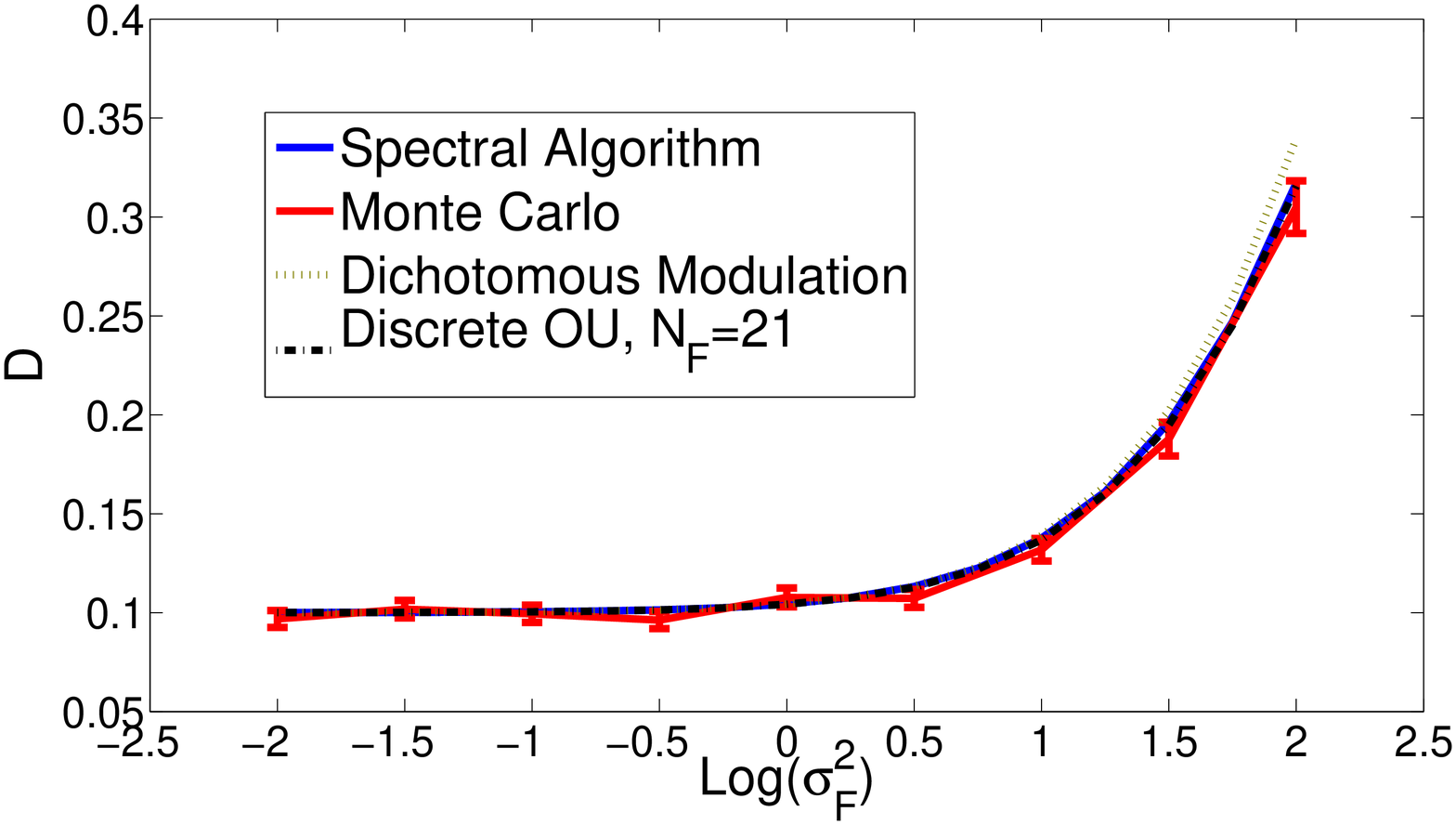}
   \label{comp1}}
 \end{subfloat}
 ~ \begin{subfloat}[$\tau=10$]{
  \centering
  \includegraphics[width=0.45\textwidth]{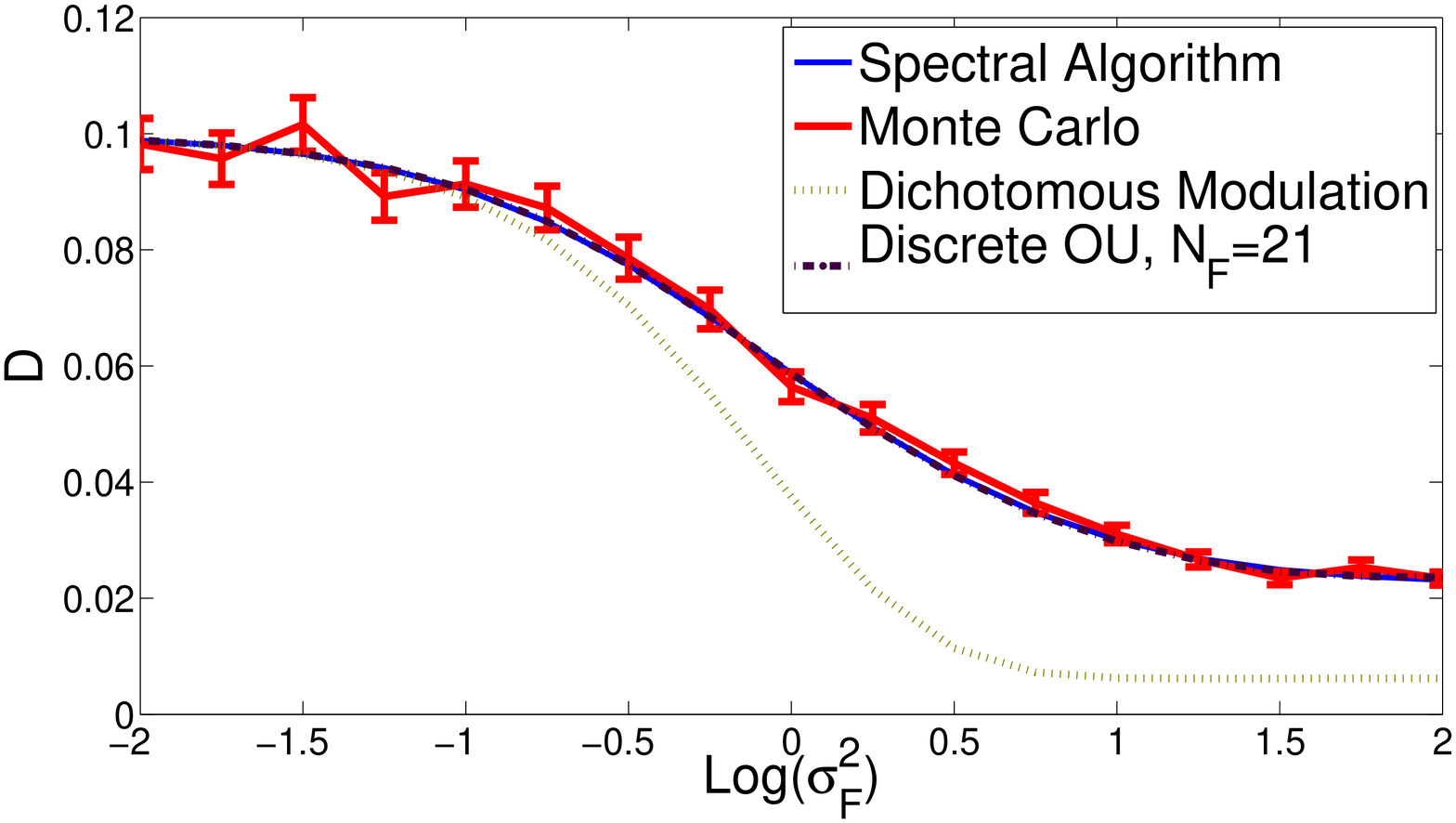}
  \label{comp2}}
 \end{subfloat} \\
 \begin{subfloat}[$\tau=0.1$]{
  \centering
  \includegraphics[width=0.45\textwidth]{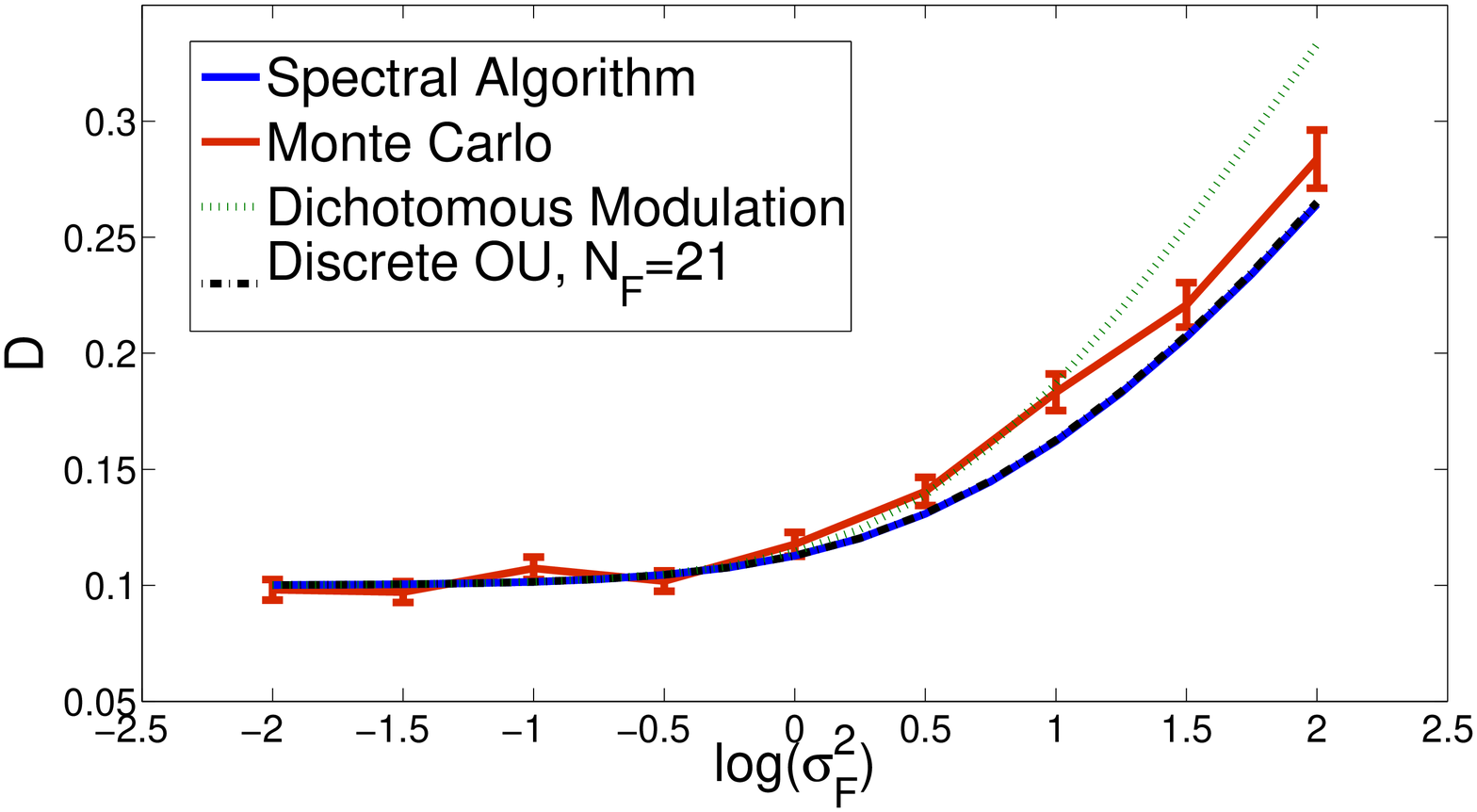}
  \label{comp3}}
 \end{subfloat}~
 \begin{subfloat}[$\tau=1$]{
  \centering
  \includegraphics[width=0.45\textwidth]{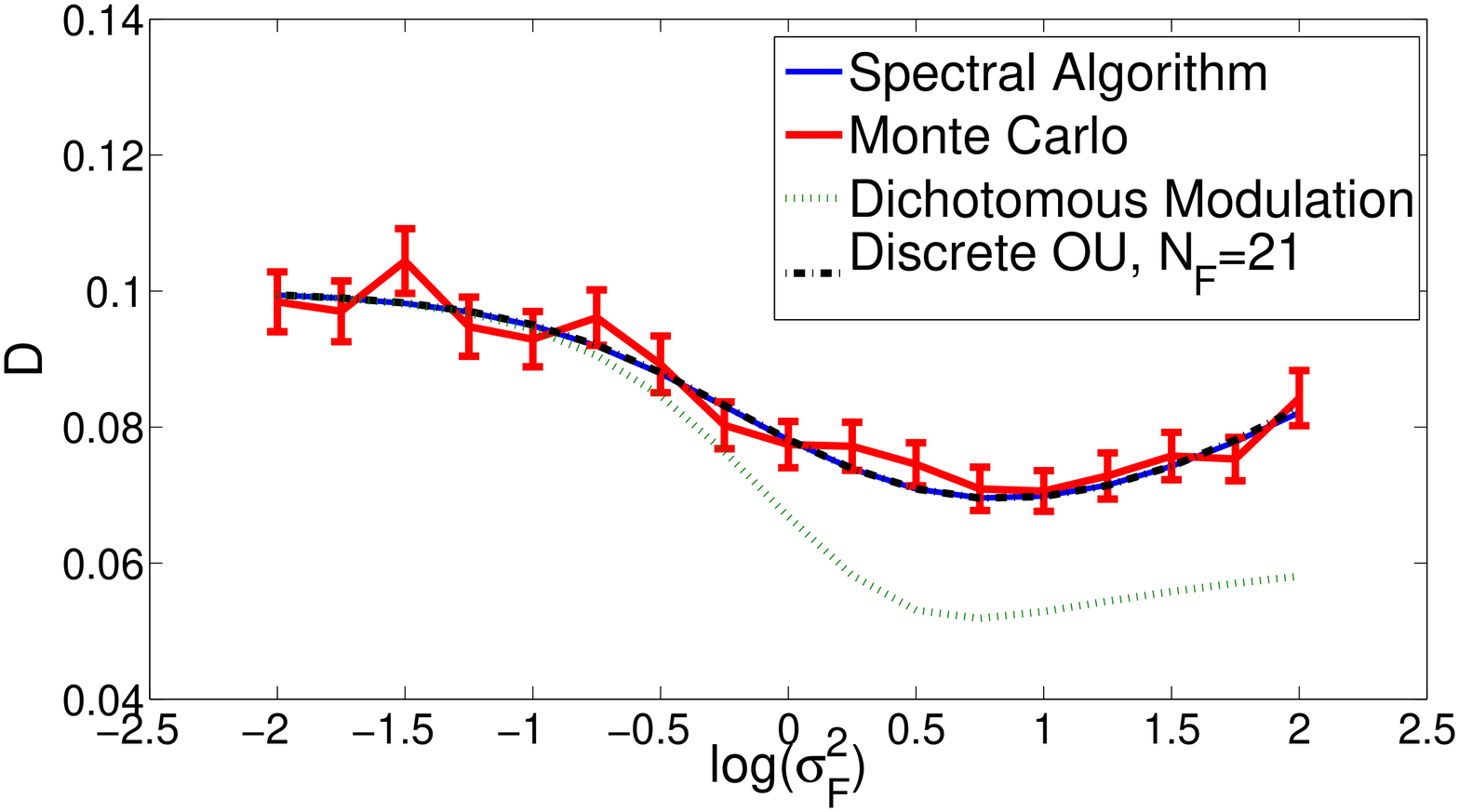}
  \label{comp4}}
 \end{subfloat}
 \caption{Effective diffusivity $\deff$ as a function of multiplicative noise variance $\sigma_F^2$ ($\theta=0.1$), computed for the OU-flashing ratchet (\ref{eq:flashingSDE}) with the spectral homogenization algorithm (solid line), the finite-volume adaptation of the WPE-based numerical algorithm ($N_F = 21$, dotted line.) The dash-dot line indicates the effective diffusivity for a flashing ratchet with dichotomous noise with same mean and correlation function as the OU-flashing ratchet. Monte Carlo simulations  after (\ref{comp1}) $3 \times 10^5$, ((\ref{comp2})-(\ref{comp4})) $10^5$ time steps (solid line with one standard deviation error bars).  } \label{fig:comparisons}

\end{figure}

We explore the performance of the homogenization algorithm and the WPE-based method for a rather simple example in which the potential is sinusoidal $\phi (x)=\- - \frac{\bar{\phi}}{2 \pi} \cos{2\pi x}$.  Due to the symmetry of the potential, the effective drift should vanish ($\ueff =0$), and we have verified (but do not show explicitly) that both methods correctly reproduce this result to the appropriate numerical accuracy.  In particular, we do not see any significant spurious drift even without designing detailed balance into the homogenization algorithm.   
In Figures \ref{comp1} and \ref{comp2} we present the results of the computations for the effective diffusivity as 
a function of the variance $\sigma_F^2$
of the modulations $ F(t) $ for two different values of the correlation time $\tau$. For $\tau=0.01$ (Figure~\ref{comp1}) we observe an enhancement of diffusivity (i.e., $\deff > \theta$) while for $\tau =10$ (Figure~\ref{comp2}) we observe a suppression of diffusivity (i.e., $\deff < \theta$). In these figures, we have used $M_s=20$ (41 Fourier coefficients) and $N_s=30$ Hermite polynomials for the spectral method, while using $M_x=500$ grid points in the $x$-direction and $N_F=21$ grid points in the $f$-direction for the WPE-based method (resulting in a Markov jump process with 21 states; see~\ref{app:ouApproximation}
for how $ \Delta f $ is chosen.) 

The Monte Carlo simulations were performed by an Euler-Maruyama discretization of the SDE (\ref{eq:flashingSDE}), with a time step $\Delta t=0.001$ and an ensemble average over $1000$ independent simulations after a large number of time steps, which varies depending on the parameters of the simulation (see the figure captions for the actual number). The diffusivity $\deff$ for the flashing ratchet with dichotomous noise was also computed via the WPE-based method, using the same number of grid points. 

We can observe from these figures how the actual number of states for the multiplicative noise plays a fundamental role as the fluctuations of $F$ become larger, especially for larger values of $\tau$.  In Figures \ref{comp3} and \ref{comp4} we present computations of the effective diffusivity $\deff$ for intermediate values of $\tau$, where this phenomenon is also observed. The parameters for the algorithm in the numerical simulations are the same as before.

We can observe from the figures that each of the methods are computing the effective diffusivity consistently for the parameter ranges explored, and that the behavior of the motor particle is sensitive to whether the flashing ratchet is discrete or continuous precisely when the correlation time is large and the amplitude of the potential modulations is not small (in our rescaled units).

For another perspective on the results, we study next the behavior of the effective diffusivity as a function of the parameters $ \tau $, $ \Diffref $, and $ \theta $, where $ \Diffref \equiv \sigma_F^2 \tau $.  This latter parameter characterizes the strength of the noise somewhat differently than the simple amplitude by also taking into account  the correlation time.  $ \Diffref $ can be thought of as a crude scaling estimate, from kinetic theory principles,  of the enhancement of the diffusivity of the motor particle due to the flashing ratchet, and should be accurate (up to constant prefactor) for the case of low Kubo number~\citep{jak:fsdpt} in which the decorrelation in the motion of the motor particle is determined essentially by the temporal decorrelation of the amplitude modulation $ F(t) $ rather than spatial decorrelation through motion across the potential landscape $ \phi (x) $.

\begin{figure}[t]
\begin{center}
\includegraphics[width=0.95\textwidth]{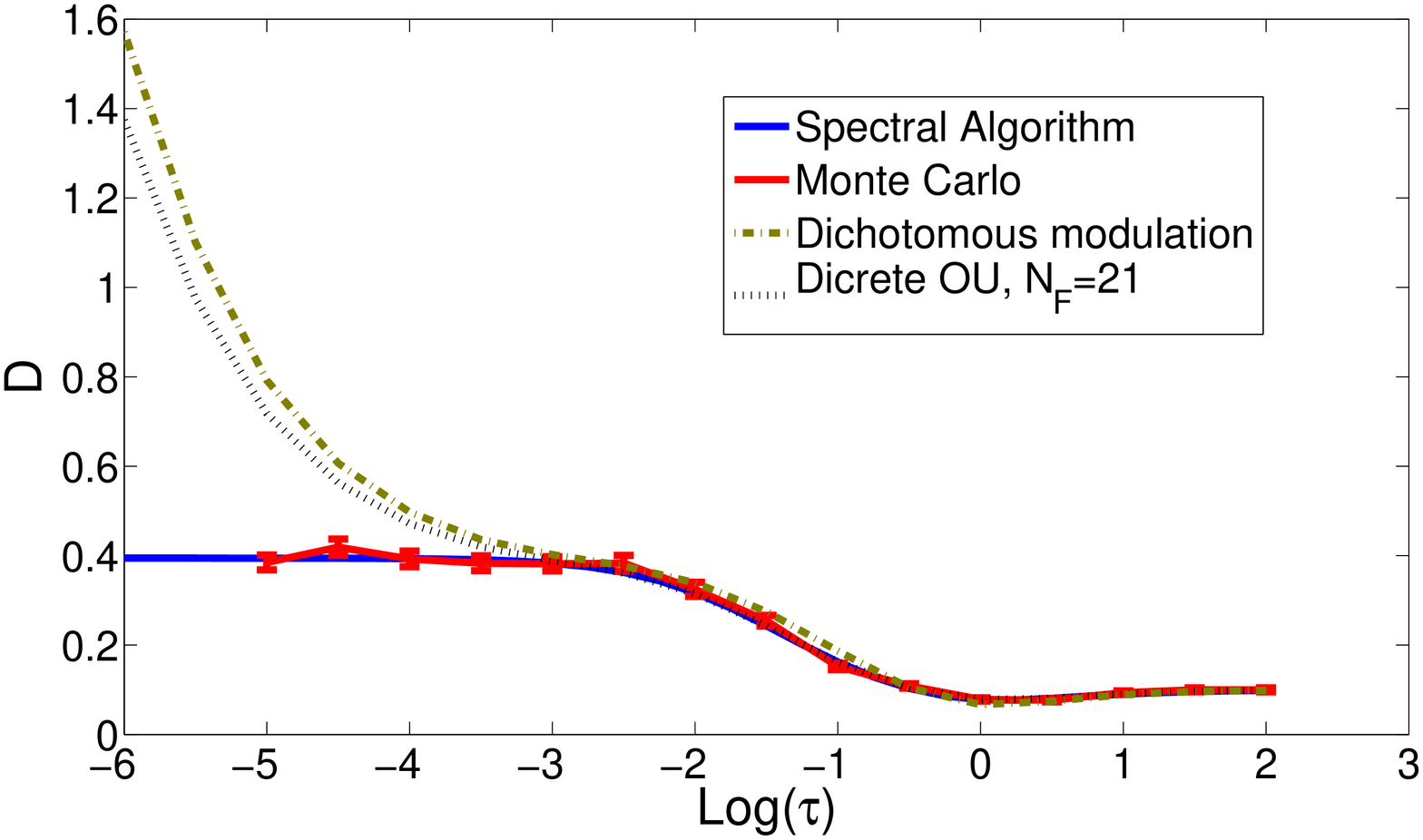}
\caption{$\deff$ as a function of $\tau$, with $\Diffref = \sigma_F^2 \tau=1$, $\theta=0.1$, computed for the OU-flashing ratchet (\ref{eq:flashingSDE}) with the spectral homogenization algorithm (solid line), the  WPE-based numerical algorithm ($N_F = 21$, dotted line), and Monte Carlo simulations after $2 \times 10^5$ time steps (solid line with one standard deviation error bars).   The dash-dot line indicates the effective diffusivity for a flashing ratchet with dichotomous noise with the same mean and correlation function as the OU-flashing ratchet.
} \label{comp5}
\end{center}
\end{figure}

In Figure \ref{comp5} we present our findings when we fix $\Diffref=1$. The parameters for the algorithm are the same as before.
We see first of all that the homogenization algorithm remains in good agreement with the Monte Carlo simulations throughout the range of correlation times presented, and that the effective diffusivity so computed agrees with the intuition described above that $ \deff \sim C \Diffref $ for small correlation time $ \tau $ (with some order unity constant $ C$ ).  On the other hand, the WPE-based algorithm with $N_F = 21$ fails to follow the Monte Carlo simulations when the correlation time $\tau$ of the Gaussian noise is very small (and consequently the noise amplitude $ \sigma_F $ is very large). In this scenario one must increase the number of states in the Markov chain approximation of the OU-process to obtain accurate results with the WPE-based algorithm.  We note also the related observation that for small correlation times $\tau $ and fixed $ \Diffref = \sigma_F^2 \tau $, the behavior of the motor particle becomes very sensitive to whether the potential modulations are continuous or discrete.  This regime corresponds to 
a limit in which $ F(t) $ approaches white noise with correlations $ \langle F(\tp) F(t+\tp) \rangle = \Diffref \delta (t) $.  
Combining the observations from Figures~\ref{fig:comparisons} and~\ref{comp5}, we see that the dichotomous Markov chain approximation to the continuous Ornstein-Uhlenbeck process for the random potential modulations creates similar behavior for the motor particle, except when the amplitude $ \sigma_F $ of the fluctuations is larger than some critical value which decreases with the correlation time $ \tau $.

\subsection{Cost Comparison between the Spectral and WPE-based numerical Methods}
\begin{figure}[!ht]
\begin{center}
\includegraphics[width=0.95\textwidth]{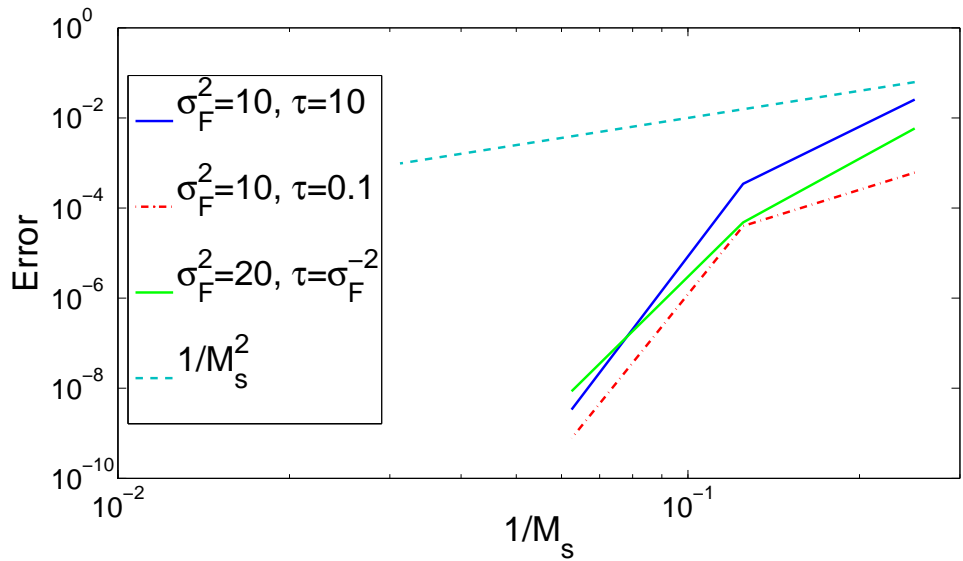}
\caption{Error in the numerical solution of $\deff$ using the spectral method as a function of the number of Fourier elements for three different choices of $\tau$. In the simulation the number of Hermite polynomials $N_s$ was kept constant at $N_s=30$.}
\label{fig:convTestSpecM}
\end{center}
\end{figure}

\begin{figure}[!ht]
\begin{center}
\includegraphics[width=0.95\textwidth]{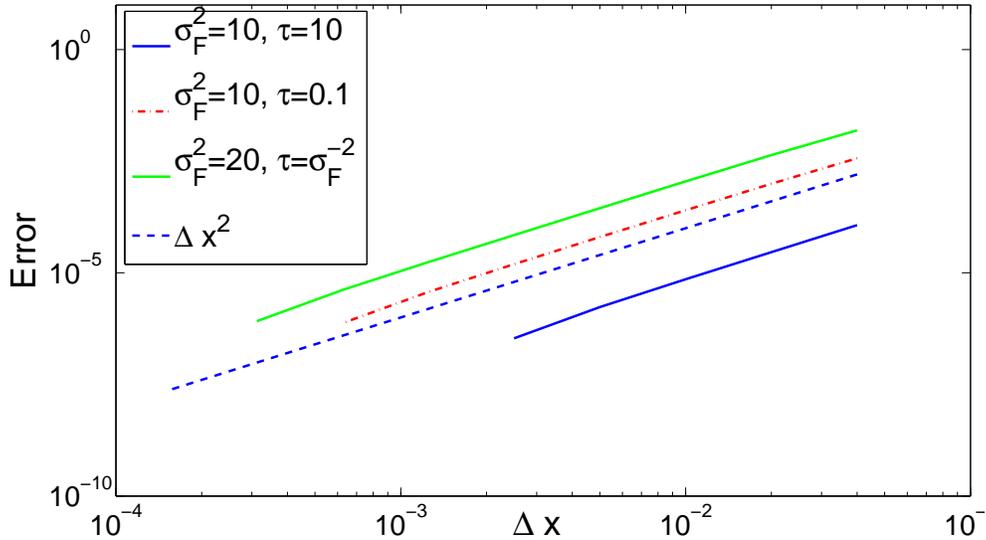}
\caption{Error in the numerical solution of $\deff$ using the WPE method as a function of the grid size $\Delta x = 1/M_x$ for three different choices of $\tau$. In the simulation the number of grid points $N_F$ in the $F$ direction was kept constant at $N_F=30$.} \label{fig:convTestWpeM}
\end{center}
\end{figure}

We present now a comparison of how the solution of the numerical methods presented above converge with respect to the number of elements taken in the approximation. As we saw in the previous section, the number of states $N_F$ in the discrete approximation of the OU process plays an important role in the accuracy of the WPE-based method, especially for large values of $\sigma_F^2$. This should come as no surprise, for the approximation is based on a finite-volume approximation of the backward-Kolmogorov equation of the OU process. Then the factor $1/N_F$ is proportional to the grid size $\Delta f$.  We present then convergence comparisons between the two methods for refinements of the discretization of the spatial variable $ x$ and the modulational noise variable $ f$.  To represent the spatial discretization, we use the number of grid elements, $M_x$, taken in the WPE approximation for the $X(t)$ process and the number of Fourier elements, $M_s$, taken in the truncation in the spectral algorithm.  For the modulational noise variable discretization, we represent the computational effort with  the number of grid points, $N_F$, taken in the finite-volume approximation of the OU process and  the number of Hermite polynomials, $N_s$, taken in the spectral algorithm. In Figure~\ref{fig:convTestSpecM} we present how the error in the numerical solution of the spectral algorithm is reduced as we increase the number of Fourier elements in the truncation while keeping the number of Hermite polynomials constant. The error is computed as usual as 
\begin{displaymath}
\mathrm{error}(M_s)=|\mathrm{D}_{M_{\mathrm{max}}}-\mathrm{D}(M_s)|,
\end{displaymath}
where $\mathrm{D}_{M_{\mathrm{max}}}$ is the solution using a large number of Fourier elements $M_{\mathrm{max}}$ (in this case it is double the number of the last simulation point), and $D(M_s)$ is the solution computed using $M_s$ Fourier terms (analogously, $M_w$ grid points in the $x$ coordinate for the WPE-based method). In Figure \ref{fig:convTestWpeM} we present the same experiment for the WPE-based method. In this case, we compute the error in the solution as we increase the number of grid points $M_x$ in the $X$ direction, while keeping fixed the number of grid points $N_F$ in the $F$-direction.

\begin{figure}[!ht]
\begin{center}
\includegraphics[width=0.95\textwidth]{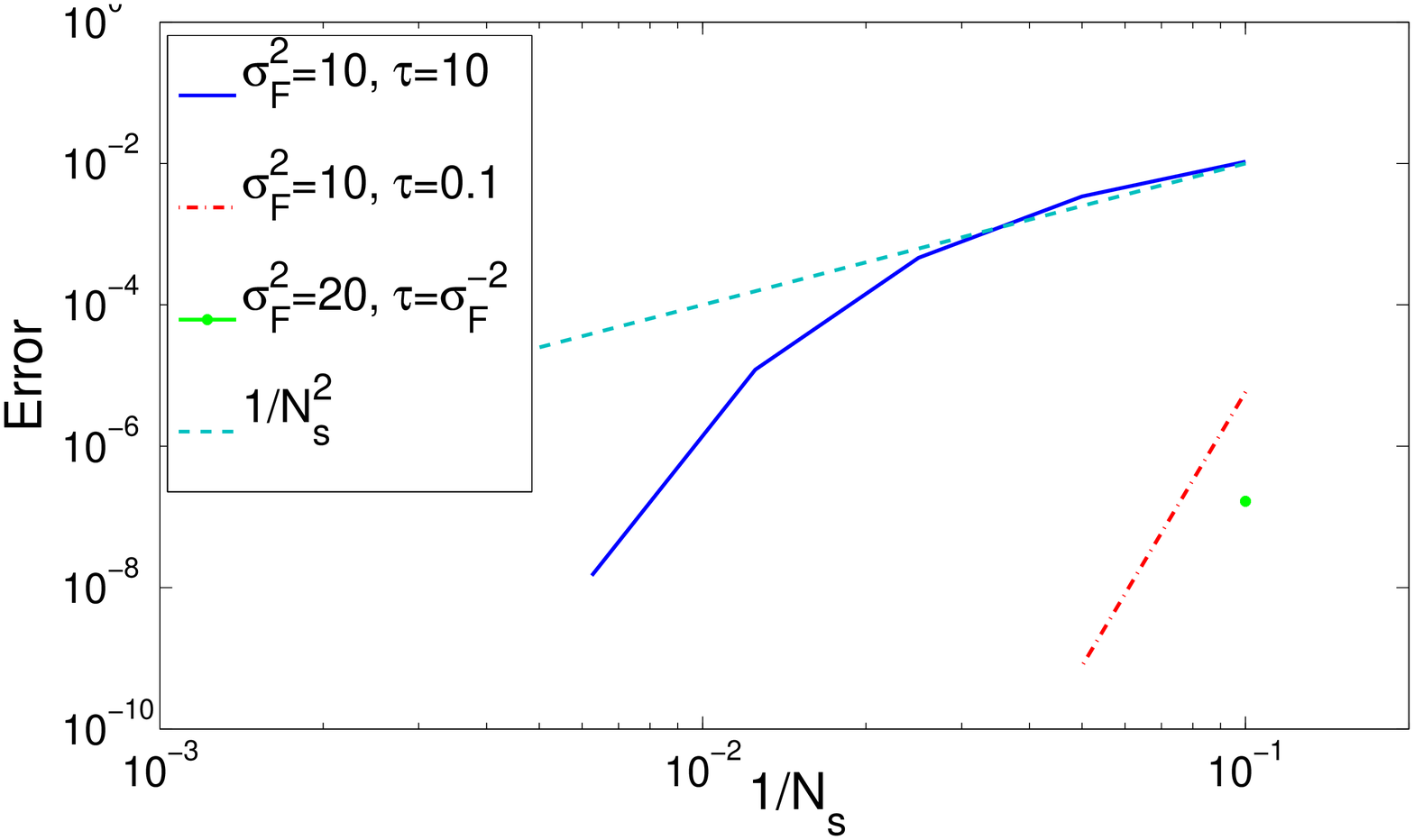}
\caption{Error in the numerical solution of $\deff$ using the spectral method as a function of the number of Hermite polynomials for three different choices of $\tau$. In the simulation the number of Fourier terms $M_s$ was kept constant at $M_s=10$.} \label{fig:convTestSpecN}
\end{center}
\end{figure}

\begin{figure}[!ht]
\begin{center}
\includegraphics[width=0.95\textwidth]{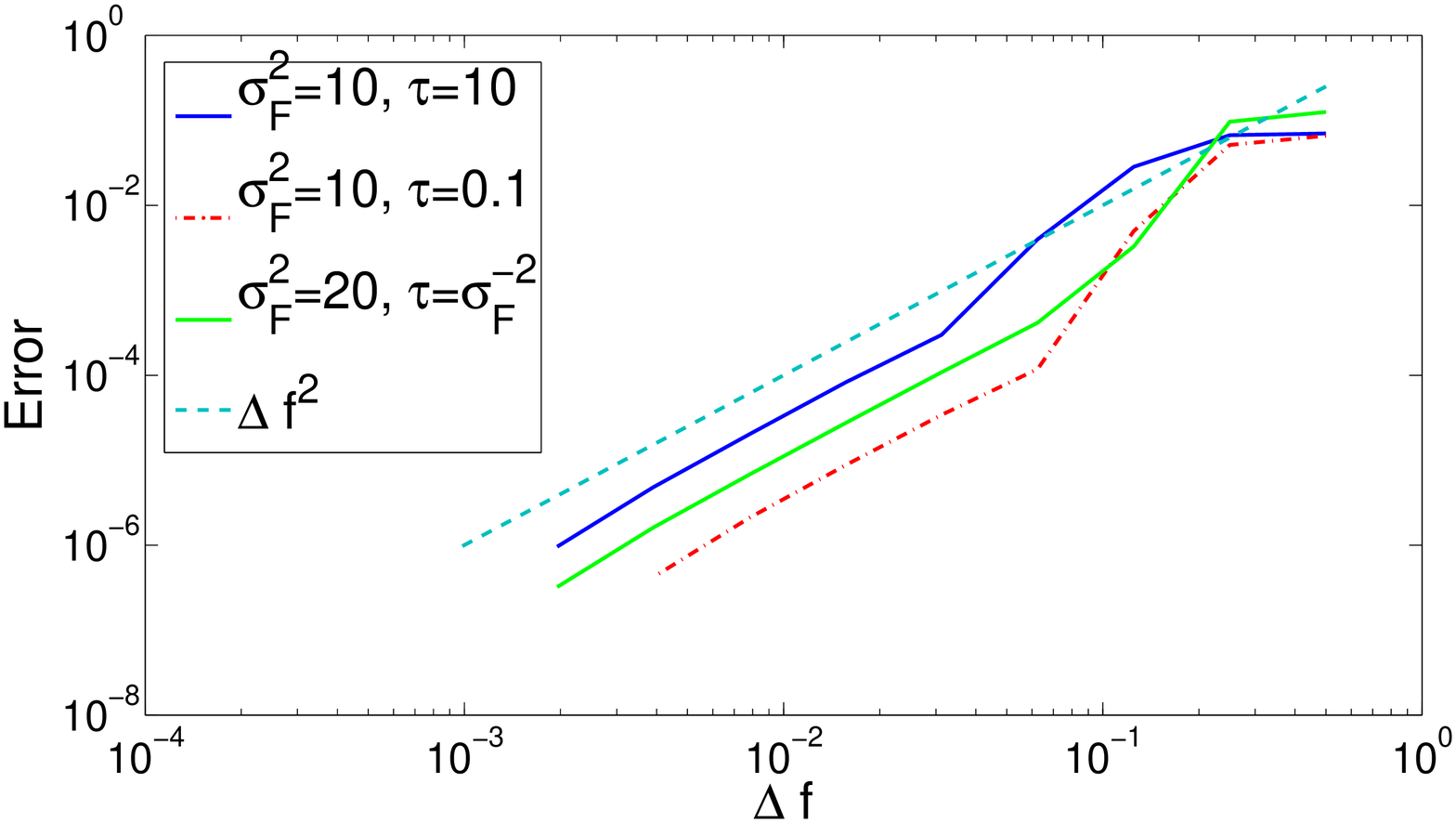}
\caption{Error in the numerical solution of $\deff$ using the WPE method as a function of the grid size $\Delta f \sim 1/N_F$ for three different choices of $\tau$. In the simulation the number of grid points $M_x$ in the $X$ direction was kept constant at $M_x=25$.} \label{fig:convTestWpeN}
\end{center}
\end{figure}

In Figures \ref{fig:convTestSpecN} we perform a similar experiment but now increasing the number of Hermite polynomials $N_s$ in the solution of the spectral method while keeping the number of Fourier elements fixed. Analogously, in Figure \ref{fig:convTestWpeN} we increase the number of grid elements $N_F$ (equivalently to decreasing the grid size $\Delta f$) while keeping the grid size $\Delta x$ constant.  

We can clearly see the $\Delta x^2$-convergence in the WPE-based method (as well as $\Delta f^2$-convergence), which is characteristic of 2nd-order finite-volume approximations. On the other hand, it is clear how  the spectral method converges faster as the number of spectral elements are increased. \\

\begin{figure}[!ht]
\begin{center}
\includegraphics[width=0.95\textwidth]{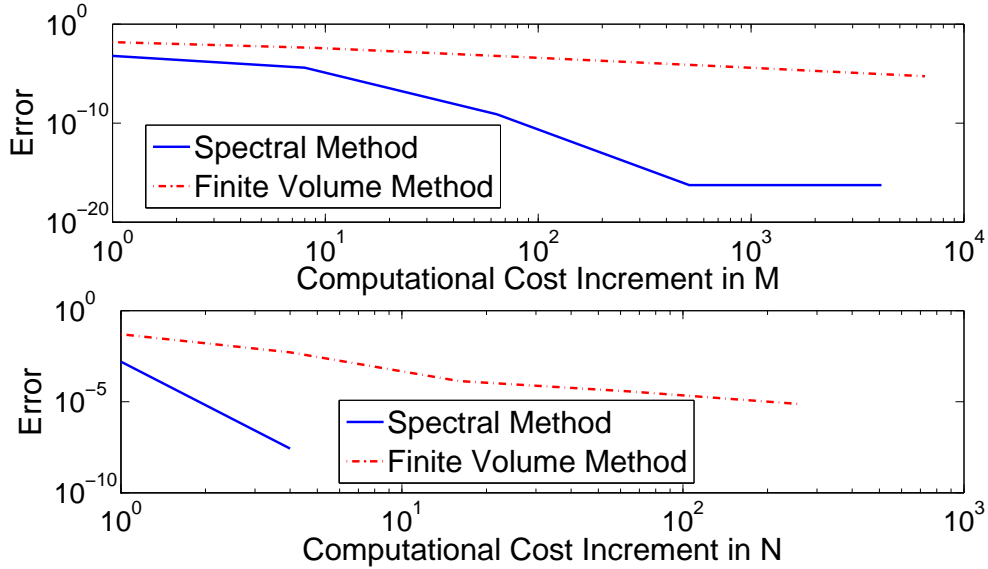}
\caption{Cost comparison between the spectral and the WPE-based (finite-volume) numerical algorithms for $\tau=0.1$, $\sigma_F^2 = 10$, and $\theta=0.1$ (case of enhanced diffusion).  Upper panel: The numerical cost of the spectral algorithm is estimated as $M^3$, while the numerical cost for the WPE-based algorithm is estimated as $M^2$. The number of $N$-elements is kept fixed at $N_s=20$ for the spectral method and $N_F=40$ for the WPE-based method. Bottom panel: The numerical cost of the spectral algorithm is estimated as $N_s$, while the numerical cost for the WPE-based algorithm is estimated as $N_F^2$. The number of $M$-elements is kept fixed at $M_s=10$ for the spectral method and $M_x=50$ for the WPE-based method. } \label{fig:costComp1}
\end{center}
\end{figure}

\begin{figure}[!ht]
\begin{center}
\includegraphics[width=0.95\textwidth]{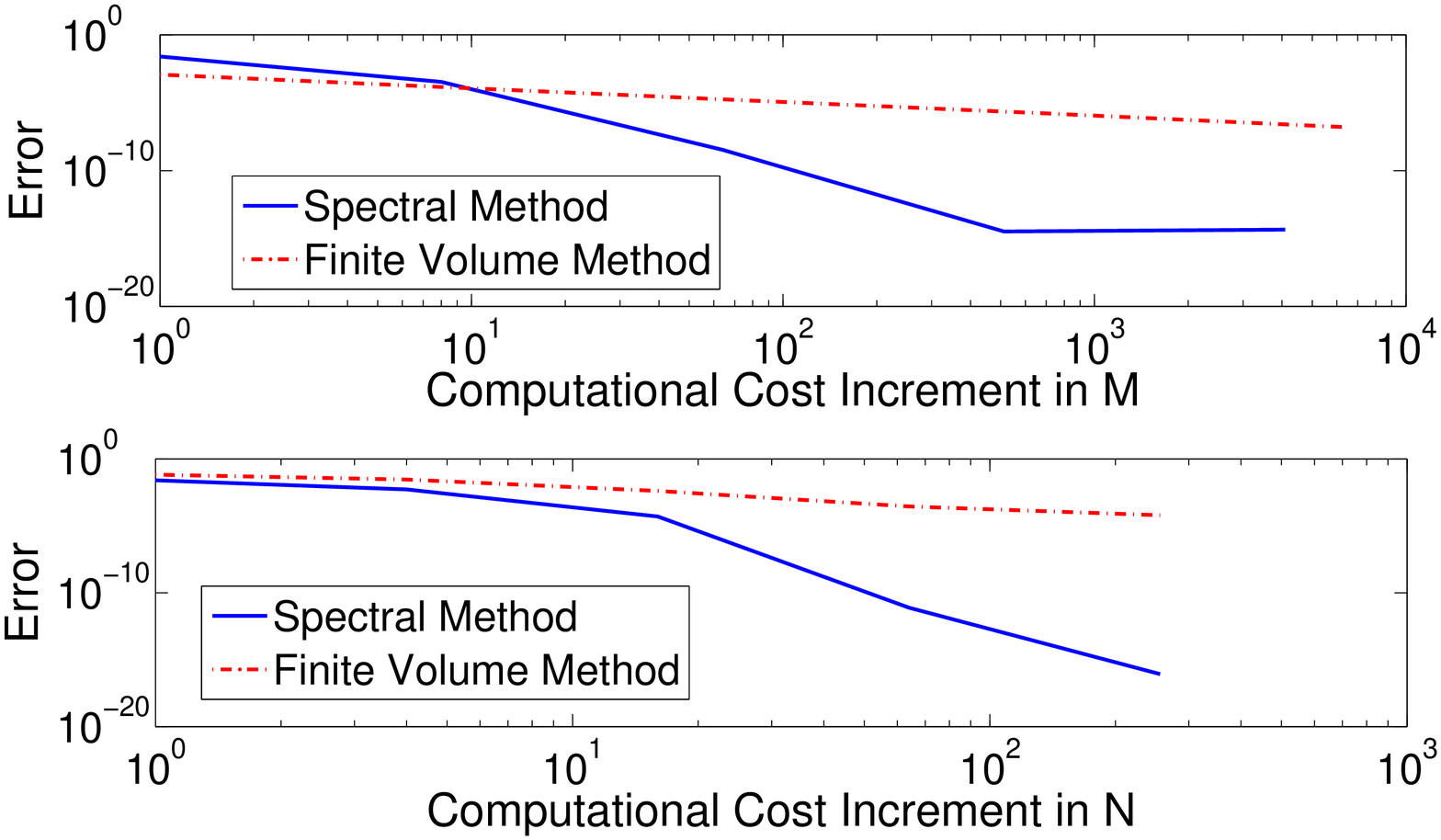}
\caption{Cost comparison between the spectral and the WPE-based (finite-volume) numerical algorithms for $\tau=10$, $\sigma_F^2=10$, and $\theta=0.1$(case of suppressed diffusion). Upper panel: The numerical cost of the spectral algorithm is estimated as $M^3$, while the numerical cost for the WPE-based algorithm is estimated as $M^2$. The number of $N$-elements is kept fixed at $N_s=20$ for the spectral method and $N_F=40$ for the WPE-based method. Bottom panel: The numerical cost of the spectral algorithm is estimated as $N_s$, while the numerical cost for the WPE-based algorithm is estimated as $N_F^2$. The number of $M$-elements is kept fixed at $M_s=10$ for the spectral method and $M_x=50$ for the WPE-based method. } \label{fig:costComp2}
\end{center}
\end{figure}

A natural question now is how the error in both methods converges as the \emph{cost} of the numerical method is increased. Although a careful analysis of the numerical cost (as given by the number of flops, for instance) is beyond the scope of this paper, we can provide a rough estimation for both methods. The WPE-based algorithm involves the solution of two $MN~\mathrm{x}~MN$ system of equations ($M$ for the grid size in $X$, $N$ for the grid size in $F$). This is performed usually in $\mathrm{O} ((MN)^3)$-flops\footnote{Using a Gauss-Seidel method, for instance.}, but 
further examination of the matrices involved in the method reveals that the system is sparse and banded, reducing the cost of the solutions to $\mathrm{O}((MN)^2)$-flops. The spectral numerical method involves the solution of two sets of recursive  systems (one for $ \rho $ and one for $ \chi $) of $N+1$ equations 
of the form (see~\ref{app:spectralMethod}),
\begin{displaymath}
 -\left(\varmatrix{Q}_n+\varmatrix{Q}_n^-\varmatrix{S}_{n+1}\right)^{-1}\varmatrix{Q}_n^+.
\end{displaymath}
Although numerically the inverse matrix is never explicitly computed, the above operation is numerically equivalent to solving $2M+1$ systems of $(2M+1)\times(2M+1)$ equations.   This operation can be done in $\mathrm{O}(M^3)$-flops, 
since all the matrices involved in this equation are also sparse and banded, so that the total cost of the spectral numerical method is $\mathrm{O}(NM^3)$-flops. The comparison between the cost of the numerical methods is done in the following way. By keeping the number of $N$-elements (either Hermite polynomials or grid points in the $F$-direction) we start with a small number of $M$-elements (both Fourier and $X$-grid points.) The number of $M$-elements is then increased such that the cost in both numerical methods is increased by (approximately) the same factor.  In other words, while we double the number of Fourier elements (increasing the cost in the spectral algorithm by a factor of eight) we triple the number of grid points (increasing the cost in the WPE-based algorithm by a factor of nine). In the same manner, keeping the number of $M$-elements fixed while we double the number of $N$-elements for the WPE-based method (increasing the cost by a factor of four), we take $4~N$ elements for the spectral algorithm (increasing the cost also by a factor of four). In Figure \ref{fig:costComp1} and Figure \ref{fig:costComp2} we show the results for two different values of $\tau$.
As we can observe from these results, the convergence of the spectral method is much faster than that of the WPE method relative to the numerical cost involved in both methods.

\section{Summary and Discussion.}
\label{sec:conc}
We have presented a novel numerical algorithm for computing the effective transport properties of the flashing ratchet with continuous Gaussian modulations (Ornstein-Uhlenbeck process). This numerical algorithm is based on a spectral decomposition of the solution to the stationary Fokker-Planck and Poisson equations that arise in homogenization theory.  The method is shown to produce results in agreement with Monte Carlo simulations, with much less computational expense. We have also compared this spectral homogenization algorithm with a finite volume variation
of another computational approach due to WPE~\citep{wang2003,hw:mcmse}, which can be applied once the continuous modulations are discretized into a continuous-time Markov chain.  Both algorithms have been shown to be theoretically equivalent, and capable of accurately reproducing the results of Monte Carlo simulations, with the error of our spectral method converging to zero more rapidly with increasing computational effort.  We have also examined to what extent the continuity or discreteness of the potential modulations affects the transport properties of the motor particle.  In one direction,  the WPE computational approach is based from the start on a discretization of the state space of the random modulations, and we have found that with 21 states, the WPE method successfully computes the effective drift and diffusivity of the flashing ratchet model over a wide range of parameters, except in the white noise limit when the correlation time of the modulations is taken small while their amplitude is taken large.  Presumably a larger number of states are needed for accurate representation by the WPE method in this regime.  From another perspective, we considered how a relatively 
crude approximation of the continuous Ornstein-Uhlenbeck process for the potential modulations in terms of a 2-state (dichotomous) Markov chain with the same mean and correlation function affects the transport properties of the flashing ratchet.  We found that the dichotomous and continuous models produced similar behavior for the motor particle over a broad range of parameters, except when either the correlation time or the amplitude of the noise is sufficiently large.  

We finally mention some directions for future exploration and development of the method presented here.   More general types of modulations, not necessarily described by Ornstein-Uhlenbeck processes, can be considered. The extension would be straightforward for modulations described by  diffusion processes for which an appropriate orthonormal basis can be constructed using the eigenfunctions of the generator of the process (e.g. the Hermite polynomials for the Ornstein-Uhlenbeck process). We believe that our algorithm can also be extended to higher dimensional problems and to systems of coupled SDEs/Fokker-Planck equations with the amplitudes of forcing terms modulated by a stochastic process. This would require the use of appropriate tensor products of Hermite polynomials and Fourier basis functions, together with appropriate preconditioning to reduce the computational cost. The rigorous numerical analysis of our algorithm, establishing convergence and analyzing its stability properties, is another natural next step.

Much of our methodology can be carried over in principle to flashing ratchet systems where the continuous-state stochastic process modulates the potential in a  more general way than its amplitude, i.e., $ \phi = \phi (x,f) $.  
The primary change in the spectral algorithm developed in Subsection~\ref{sec:spec} is that the $f$ dependence of $ \phi $ would need to be expanded with respect to the Hermite polynomials (or other basis appropriate to the generator of the continuous-state  process $ F(t) $), and this would introduce more coupling between spectral coefficients of the desired solutions.  That is, the spectral method should enjoy comparable complexity to the results presented here for the amplitude-modulated flashing ratchet~(\ref{gauss_ratchet}) provided $ \phi (x,f) $ can be well approximated by a low order expansion with respect to Hermite polynomials (or other appropriate basis functions).  For molecular motors models that include both discrete-state and continuous-state stochastic processes, one could contemplate a hybrid approach between the WPE algorithm and the spectral approach presented in Subsection~\ref{sec:spec}.  More precisely, we would advocate use of a spectral approach in handling the continuous state stochastic processes modulating the molecular motor model, while the discrete-state components could be treated with the WPE method~\citep{wang2003,hw:mcmse}.   In particular, the spatial variable could be discretized in the WPE manner, respecting detailed balance in thermal equilibrium, while the non-equilibrium continuous stochastic driving components are handled spectrally.   This is possible because the discretization of the spatial variable into Fourier modes  in our spectral numerical method in Subsection~\ref{sec:spec} was done for methodological coherence (i.e., a spectral representation jointly in space and modulational noise), but this was neither necessary nor fundamental to our approach.  Therefore, one could in principle combine a spectral representation for the continuous stochastic modulation with not only a WPE discretization of the spatial variable, but with more recent variations~\citep{latorre2011,WangOster:2005}  which feature some technical improvements.  Such integration of methods would require some care and thought, particularly since the spatial variable should presumably be discretized in one consistent manner, but the underlying unity of the homogenization and WPE frameworks shown in Section~\ref{sec:eqwpehom} indicates the likely intellectual coherence of such a hybrid approach.  

We have, based on the considerations of~\citet{elston2000}, focused our attention on the computation of the effective drift and diffusivity of Brownian motor models.  In principle, trajectories of the Brownian motor could also be simulated through a spectral Karhunen-Lo\'{e}ve decomposition of the continuous stochastic modulation $ F(t) $ \emph{in path space} rather than in terms of its probability density (measure)~\citep{kloeden1992kl}.  Such an approach would not be directly connected to the homogenization procedure discussed in Section~\ref{sec:hom}, nor do we see any particular benefit to such a simulation procedure relative to more straightforward approaches.  Our advocacy for the homogenization procedure presented in Section~\ref{sec:hom} is precisely for 
providing a flexible framework for the purpose of efficiently computing effective transport coefficients of Brownian motor models, which first of all can be smoothly connected with existing successful approaches such as the WPE method, and can moreover be flexibly discretized such as in the spectral manner described in Subsection~\ref{sec:spec} to more efficiently handle continuous-state stochastic modulation components.

\appendix
\section{The Spectral Numerical Method} \label{app:spectralMethod}
For simplicity in the presentation we choose $\phi(x)$ to be $\phi(x)=-\frac{\bar{\phi}}{\omega_1} \cos{\omega_1 x}$, with $\omega_1=2\pi$, although more complex potentials may be considered.  The spectral representation of $\phip (x)$ is then simply,
\begin{displaymath}
\phip (x) = \frac{1}{2\mathrm{i}}\left(\e{\mathrm{i}\omega_1 x}-\e{\mathrm{i}\omega_{-1}x}\right).
\end{displaymath}

This leads to the following spectral representation of (\ref{eq:system_pi}) in terms of the Fourier coefficients $ \{\pi_n^j\} $ from Eq.~\eqref{eq:pifour}, with $j=-\infty,\ldots,\infty$,
\begin{subequations}
 \begin{equation}
 \sigma_F\frac{1}{2}\omega_j \left(\pi_1^{j-1}-\pi_1^{j+1}\right) - \theta \omega_j^2\pi_0^j =0
 \end{equation}
\begin{equation} \label{eq:pi_nj}
\varmatrix{L}^-_{n} \boldsymbol{\pi}_{n+1}+ \varmatrix{L}_{n} \boldsymbol{\pi}_{n}+ \varmatrix{L}^+_{n}\boldsymbol{\pi}_{n-1}=0, \quad n=1,2,\ldots
\end{equation}
\end{subequations}
where $ \boldsymbol{\pi}_n $ is an infinite column vector of Fourier coefficients of $ \pi_n (x) $, and the matrix-vector products above are shorthand for the following operations on Fourier coefficients:
\begin{equation}
\begin{array}{rl}
\left[\varmatrix{L}^-_n \boldsymbol{\pi}_{n+1}\right]^j=&\ds \sqrt{(n+1)} \sigma_F  \frac{1}{2}\omega_j \left(\pi_{n+1}^{j-1}-\pi_{n+1}^{j+1} \right),\\
~\\
\left[\varmatrix{L}^+_n\boldsymbol{\pi}_{n-1}\right]^j=&\ds \sqrt{n} \sigma_F \frac{1}{2}\omega_j \left(\pi_{n-1}^{j-1}-\pi_{n-1}^{j+1} \right),\\
~\\
\left[\varmatrix{L}_n\boldsymbol{\pi}_{n}\right]^j=&\ds \left(-\theta \omega_j^2-n\tau^{-1}\right)  \pi_{n}^{j}.
\end{array}
\label{eq:lmatdefs}
\end{equation}
The normalization of the solution, $\rho(x,f)$ implies furthermore,
\begin{eqnarray*}
 \int_0^1 \int_{-\infty}^\infty \rho(x,f) \de f \dx &=&  \int_0^1 \int_{-\infty}^\infty \rho_F(f) \sum_{n=0}^\infty \pi_n(x) H_n(f) \de f \dx \\
&=&   \int_0^1 \sum_{n=0}^\infty \pi_n(x) \delta_{n,0} \dx \\
&=& \int_0^1 \sum_j \pi_0^j \e{\mathrm{i}\omega_j x} \dx \\
&=& \pi_0^0 =1.
\end{eqnarray*}
It can also be noticed from the $j=0$ component of equations (\ref{eq:pi_nj}) that $\pi_n^0=0$, $n=1,2,\ldots$. We now approximate $\pi(x)$ by applying a Galerkin truncation to the infinite series at suitable finite values $ N_s $ and $ M_s$,
\begin{displaymath}
 \pi(x) \approx \sum_{n=0}^{N_s} \sum_{j=-M_s}^{M_s} \, \pi_n^j  \e{\mathrm{i}\omega_j x} H_n(f).
\end{displaymath}
By taking the  values of $ \pi_n^0 $ as given above, 
the system of equations~(\ref{eq:system_pi}) becomes then a finite system of $(2M_s)\times(N_s+1)$ linear equations, which can be written as, 
\begin{subequations} \label{eq:system_pinj}
\begin{eqnarray}
\varmatrix{Q}^-_0 \boldsymbol{\pi}_1+\varmatrix{Q}_0 \boldsymbol{\pi}_0&=& \boldsymbol{0}, \\
\varmatrix{Q}^-_1 \boldsymbol{\pi}_2 + \varmatrix{Q}_1 \boldsymbol{\pi}_1+\varmatrix{Q}^+_1 \boldsymbol{\pi}_0&=&\varmatrix{B}_0, \notag \\
 \vdots & & \vdots \notag \\
  \varmatrix{Q}_{N_s} \boldsymbol{\pi}_{N_s}+\varmatrix{Q}^+_{N_s} \boldsymbol{\pi}_{N_s-1}&=&\boldsymbol{0}. \notag
\end{eqnarray}
\end{subequations}

The matrices $\varmatrix{Q} = \{Q_{l+M_s+1,j+M_s+1}\}, j,l = -M_s,\ldots,-1,1,\ldots,M_s$ then take the form,
\begin{subequations}
\begin{equation}
 \left[\varmatrix{Q}^-_0\right]_{l+M_s+1,j+M_s+1}=\left\{ \begin{array}{l l}
                                      \ds  - \sigma_F \frac{1}{2}\omega_l		 & \textrm{if } j=l+1, \, l=-M_s,\ldots,-2,1,\ldots, M_s-1, \vspace{10pt} \\
                                       \ds \sigma_F \frac{1}{2}\omega_l & \textrm{if } j=l-1, \, l=-M_s+1,\ldots,-1,2,\ldots,M_s \vspace{10pt} \\
					0 & \textrm{otherwise}.
                                      \end{array} \right.
\end{equation}
\begin{equation}
 \left[\varmatrix{Q}_0\right]_{l+M_s+1,j+M_s+1}=\left\{ \begin{array}{l l}
                                      \ds  -\theta\omega_l^2		 & \textrm{if } j=l, \, l=-M_s,\ldots,-1,1,\ldots, M_s, \vspace{10pt} \\
					0 & \textrm{otherwise}.
                                      \end{array} \right.
\end{equation}
\begin{displaymath}
 \left[\varmatrix{B}_0\right]_{l+M_s+1}=\sigma_F \frac{1}{2}\omega_{-1} \delta_{l,-1} - \sigma_F \frac{1}{2}\omega_{1} \delta_{l,1},
\end{displaymath}
and for $n=1,2,\ldots,N_s$,
\begin{equation}
 \left[\varmatrix{Q}^-_n\right]_{l+M_s+1,j+M_s+1}=\left\{ \begin{array}{l l}
                                      \ds  - \sigma_F \sqrt{n+1}\frac{1}{2}\omega_l		 & \textrm{if } j=l+1, \, l=-M_s,\ldots,-2,1,\ldots, M_s-1, \vspace{10pt} \\
                                       \ds \sigma_F\sqrt{n+1}\frac{1}{2}\omega_l & \textrm{if } j=l-1, \, l=-M_s+1,\ldots,-1,2,\ldots,M_s \vspace{10pt} \\
					0 & \textrm{otherwise}.
                                      \end{array} \right.
\end{equation}

\begin{equation}
 \left[\varmatrix{Q}^+_n\right]_{l+M_s+1,j+M_s+1}=\left\{ \begin{array}{l l}
                                      \ds  - \sigma_F  \sqrt{n}\frac{1}{2}\omega_l		 & \textrm{if } j=l+1, \, l=-M_s,\ldots,-2,1,\ldots, M_s-1, \vspace{10pt} \\
                                       \ds \sigma_F \sqrt{n }\frac{1}{2}\omega_l & \textrm{if } j=l-1, \, l=-M_s+1,\ldots,-1,2,\ldots,M_s \vspace{10pt} \\
					0 & \textrm{otherwise}.
                                      \end{array} \right.
\end{equation}

\begin{equation}
 \left[\varmatrix{Q}_n\right]_{l+M_s+1,j+M_s+1}=\left\{ \begin{array}{l l}
                                      \ds  -\theta\omega_l^2-n\tau^{-1}		 & \textrm{if } j=l, \, l=-M_s,\ldots,-1,1,\ldots, M_s, \vspace{10pt} \\
					0 & \textrm{otherwise}.
                                      \end{array} \right.
\end{equation}
\label{eq:qmatdefs}
\end{subequations}

This system is then solved recursively for $\boldsymbol{\pi}_{N_s-1}, \ldots, \boldsymbol{\pi}_2$ in the form
\begin{subequations}
\begin{equation}
\boldsymbol{\pi}_n = \varmatrix{S}_n \boldsymbol{\pi}_{n-1} , n=N_s,\ldots,1,
\end{equation}
 with
\begin{eqnarray}
 \varmatrix{S}_{N_s} &=& \varmatrix{Q}_{N_s}^{-1} \varmatrix{Q}_{N_s}^{+}, \\
 \varmatrix{S}_n &=& - (\varmatrix{Q}_n + \varmatrix{Q}_n^{-} \varmatrix{S}_{n+1})^{-1} \varmatrix{Q}_n^{+},  n = N_s-1,\ldots,2.  
 \end{eqnarray}
 \label{eq:downrecurse}
 \end{subequations}
This leaves us with the following set of equations:
\begin{displaymath}
\begin{array}{rl}
\varmatrix{Q}^-_0 \boldsymbol{\pi}_1+\ds \varmatrix{Q}_0 \boldsymbol{\pi}_0=&0,\\
~\\
\ds \left( \varmatrix{Q}^-_1\varmatrix{S}_2 + \varmatrix{Q}_1 \right) \boldsymbol{\pi}_1+\varmatrix{Q}^+_1 \boldsymbol{\pi}_0=&\varmatrix{B}_0.
\end{array}
\end{displaymath}
which can be solved for $\boldsymbol{\pi}_0$ and $\boldsymbol{\pi}_1$. 
Then $\boldsymbol{\pi}_n$, $n=2,3,\ldots$ can then be recovered using Eq.~(\ref{eq:downrecurse}).
$\ueff$ is then computed as follows. From (\ref{u_c4}),
\begin{equation}
\begin{array}{rl}
\ds \ueff&=\ds -\int_0^1 \int_{-\infty}^\infty \, \phip (x)f\rho(x,f) \, \de f \, \dx \vspace{7pt} \\
&=\ds -\int_0^1\int_{-\infty}^\infty \, \phip (x)f \, \rho_{f}(f)\sum_{n=0}^\infty \, \pi_n(x)H_n(f) \, \de f\, \de x \vspace{7pt} \\
&= \ds  -\sum_{n=0}^\infty \, \int_0^1 \phip (x) \pi_n(x)  \, \int_{-\infty}^\infty \, \rho_{f}(f)fH_n(f) \, \de f\, \de x \vspace{7pt} \\
&= \ds -\sum_{n=0}^\infty \, \int_0^1 \phip (x) \pi_n(x) \sigma_F \delta_{1,n} \de x \vspace{7pt} \\
&= \ds -\sigma_F \int_0^1 \phip (x) \pi_1(x) \dx \vspace{7pt} \\
&=\ds - \sigma_F \int_0^1 \frac{1}{2\mathrm{i}} \left(\e{\imi \omega_1 x} - \e{\imi\omega_{-1} x}\right) \sum_j \pi_1^j\e{\imi \omega_j x} \vspace{7pt} \\
&=\ds - \frac{\sigma_F}{2\imi} \left(\pi_1^{-1} - \pi_1^1\right) \vspace{7pt} \\
&=\sigma_F  \mathrm{Im} \pi_1^1,
\end{array}
\label{eq:uspec}
\end{equation}
since $\ds \pi_n^{-j}=\ds \overline{\pi_n^j}$, where $\overline{A}$ is the complex conjugate of $A \in \mathbb{C}$. \\

The solution to the cell problem (\ref{cell_c4}) to compute $\deff$ is done similarly. We express $\chi(x,f)$ in its Hermite polynomial decomposition,
\begin{displaymath}
 \chi(x,f)=\sum_{n=0}^\infty \chi_n(x) H_n(f).
\end{displaymath}
Upon substitution of the above expression in (\ref{cell_c4}) and using the orthogonality of $H_n$, we obtain the following set of equations,
\begin{subequations}  \label{eq:chi_n}
\begin{equation}
 \phip (x) \sigma_F \partial_x \chi_1 - \theta \partial_{xx}\chi_0 =\ueff, 
\end{equation}
\begin{equation}
 \phip (x) \sqrt{2} \sigma_F \partial_x \chi_2 -\theta \partial_{xx}\chi_1+\tau^{-1}\chi_1+  \phip (x) \sigma_F \partial_x \chi_0=-\sigma_F \phip,
\end{equation}
\begin{equation}
-( \mathcal{L}^-)^{\transp}_n\chi_{n+1}-\mathcal{L}^{\transp}_n\chi_n- (\mathcal{L}^+)^{\transp}_n \chi_{n-1}=0 , \quad n=2,3,\ldots  
\end{equation}
\end{subequations}
where the operators $ \mathcal{L}^{-}$,$\mathcal{L}$, and $ \mathcal{L}^{+} $ are defined in Eq.~(\ref{eq:lopdefs}), and $ \transp $ denotes the adjoint.
By decomposing $\chi_n(x)$ in terms of its Fourier series,
\begin{displaymath}
 \chi_n(x)=\sum_{j=-\infty}^\infty \chi_n^j \e{\imi \omega_jx},
\end{displaymath}
the set of equations (\ref{eq:chi_n}) becomes the linear system, for $j=-\infty,\ldots,\infty$,
\begin{eqnarray*}
\sigma_F \frac{1}{2}\left( \omega_{j-1}\chi_1^{j-1} - \omega_{j+1}\chi_1^{j+1} \right) + \theta \omega_j^2 \chi_0^j &=& -\ueff \delta_{j,0}, \\
\left[ (\varmatrix{L}^-)^{\transp}_1\boldsymbol{\chi}_{2}+\varmatrix{L}^{\transp}_1\boldsymbol{\chi}_1+ (\varmatrix{L}^+)^{\transp}_1 \boldsymbol{\chi}_{0}\right]^{j}&=&\sigma_F \frac{1}{2\imi} \delta_{j,-1}-\sigma_F \frac{1}{2\imi} \delta_{j,1},\\
 (\varmatrix{L}^-)^{\transp}_n\boldsymbol{\chi}_{n+1}+\varmatrix{L}^{\transp}_n\boldsymbol{\chi}_n+ (\varmatrix{L}^+)^{\transp}_n \boldsymbol{\chi}_{n-1}&=&\boldsymbol{0} , \quad n=2,3,\ldots  
\end{eqnarray*}
where $ \boldsymbol{\chi}_n $ denotes the sequence of Fourier coefficients of $ \chi_n (x) $, and $ \varmatrix{L}^-$,$\varmatrix{L}$, and $ \varmatrix{L}^{+} $ are defined in Eq.~(\ref{eq:lmatdefs}). 
In order to solve this infinite set of equations, we apply a Galerkin truncation to the spectral decomposition of $\chi(x,f)$ at some appropriate values $M_s $ and $ N_s$:
\begin{displaymath}
 \chi(x,f) \approx \sum _{n=0}^{N_s} \sum_{j=-M_s}^{M_s} \chi_n^j \e{\imi \omega_j x}H_n(f),
\end{displaymath}
so we get the finite $(2M_s+1)(N_s+1)$ set of algebraic equations,
\begin{subequations}
\begin{eqnarray}
(\varmatrix{Q}^-_0)^{\transp} \boldsymbol{\chi}_1 +\varmatrix{Q}_0^{\transp} \boldsymbol{\chi}_0 &=&\boldsymbol{B}_0, \\
( \varmatrix{Q}^-_1)^{\transp} \boldsymbol{\chi}_2 +\varmatrix{Q}_1^{\transp} \boldsymbol{\chi}_1+(\varmatrix{Q}^+)^{\transp}_1\boldsymbol{\chi}_0 &=&\boldsymbol{B}_1, \\
 (\varmatrix{Q}^-_n)^{\transp} \boldsymbol{\chi}_{n+1} +\varmatrix{Q}_n^{\transp} \boldsymbol{\chi}_n+(\varmatrix{Q}^+_n)^{\transp} \boldsymbol{\chi}_{n-1} &=&\boldsymbol{0}, \quad n=2,3,\ldots,N_s,
\end{eqnarray}
\label{eq:qeqs}
\end{subequations}
where the matrices $ \varmatrix{Q}^-_n$, $\varmatrix{Q}_n$, and $ \varmatrix{Q}^+_n $ were defined in (\ref{eq:qmatdefs}), and 
\begin{displaymath}
 \left[\boldsymbol{B}_0\right]_{l+M_s+1}=-\ueff \delta_{l,0},
\end{displaymath}
\begin{displaymath}
 \left[\boldsymbol{B}_1\right]_{l+M_s+1}=\sigma_F \frac{1}{2\imi} \delta_{l,-1} - \sigma_F \frac{1}{2\imi} \delta_{l,1}.
\end{displaymath}
The supermatrix implicitly defined by the left hand side of this system does have a zero eigenvalue, with right eigenvector  $ (\boldsymbol{\delta}_0,\boldsymbol{0},\ldots,\boldsymbol{0})^{\transp} $ and left eigenvector $ (\boldsymbol{\pi}_0,\boldsymbol{\pi}_1,\ldots,\boldsymbol{\pi}_{N_s}) $, all inherited from discretization of the operator $ \mathcal{L} $.  
Solvability of (\ref{eq:qeqs}) follows from verifying that the supervector composed of the right hand sides is indeed orthogonal to the null left eigenvector of the supermatrix, i.e.,
\begin{equation*}
\boldsymbol{\pi}_0 \cdot \boldsymbol{B}_0 + \boldsymbol{\pi}_1 \cdot \boldsymbol{B}_1 = 0
\end{equation*}
as follows from the definitions of these vectors and the formula (\ref{eq:uspec}) for the effective velocity $ \ueff $.  A unique solution is obtained by imposing $ \chi_0^0 = 0 $, the analogue of the constraint (\ref{eq:chizeroadd}) for the continuous formulation.

The system of equations is then solved similarly as we did for $\pi_n(x)$, beginning with writing:
\begin{subequations}
\begin{eqnarray}
\boldsymbol{\chi}_n &= & \varmatrix{Z}_n \boldsymbol{\chi}_{n-1}, \quad n=N_s,N_s-1,\ldots,2,  \label{eq:chin}\\
\varmatrix{Z}_{N_s} &=& - (\varmatrix{Q}_{N_s}^+ \varmatrix{Q}_{N_s}^{-1})^{\transp}, \\
\varmatrix{Z}_n &=& - \left[\varmatrix{Q}_n^+ (\varmatrix{Z}_{n+1}^{\transp}
\varmatrix{Q}_n^{-}  + \varmatrix{Q}_n)^{-1}\right]^{\transp}, n=N_s-1,\ldots,2. \label{eq:zn}
\end{eqnarray}
\end{subequations}

From the equation for $n=1$, we can write,
\begin{equation}
 \boldsymbol{\chi}_1=\tilde{\boldsymbol{B}_1}+\varmatrix{Z}_1\boldsymbol{\chi}_0, \label{eq:chia}
\end{equation}
where $ \varmatrix{Z}_1 $ is defined by the same formula as for $ n \geq 2 $ in Eq.~(\ref{eq:zn}), and
\begin{displaymath}
 \tilde{\boldsymbol{B}_1}=\left( \varmatrix{P}^-_1 \varmatrix{Z}_2+\varmatrix{P}_1\right)^{-1} \boldsymbol{B}_1,
\end{displaymath}
Substituting this expression in the equation for $n=0$ yields
\begin{equation} \label{eq:chi_0}
\varmatrix{Z}_0 \boldsymbol{\chi}_0 = \tilde{\boldsymbol{B}_0},
\end{equation}
with
\begin{displaymath}
\varmatrix{Z}_0= \varmatrix{P}^-_0 Z_1+\varmatrix{P}_0,
\end{displaymath}
and
\begin{displaymath}
\tilde{\varmatrix{B}_0}= \varmatrix{B}_0-\varmatrix{P}^-_0 \tilde{\varmatrix{B}_1}.
\end{displaymath}
We handle the degeneracy of the matrix $ \varmatrix{Z}_0 $ by imposing $ \chi_0^0 = 0 $ and solving for the remaining components of $ \boldsymbol{\chi}_0 $.  The solution is completed by computing $ \{\boldsymbol{\chi}_{n}\}_{n=1}^{N_s} $ through Eqs.~(\ref{eq:chia}) and~(\ref{eq:chin}).  

From Eq.~(\ref{eq:deff}), a similar calculation to that in Eq.~(\ref{eq:uspec}) yields 
\begin{eqnarray} \label{eq:deff_spectral}
\deff&=&\theta+\imi \frac{ \sigma_F }{2} \sum_{n=0}^{N_s} \sqrt{n+1} \left[\sum_{j=-M_s}^{M_s} \chi_{n+1}^j \overline{\pi_n^{j+1}} - \chi_{n+1}^j \overline{\pi_n^{j-1}} + \chi_{n}^j \overline{\pi_{n+1}^{j+1}}-\chi_{n}^j \overline{\pi_{n+1}^{j-1}}\right] \notag \\
& &+4\pi \theta \imi \sum_{n=0}^{N_s} \sum_{j=-M_s}^{M_s} j \chi_n^j \overline{\pi_n^j}.
\end{eqnarray}

\section{Discrete-State Approximation of the Ornstein-Uhlenbeck Process} \label{app:ouApproximation}
In our numerical experiments, we discretize the state space for the noise variable with $ N_F = 2 n_F + 1 $ states, equally spaced with interval $ \Delta f $, and centered about $0 $: 
$\Smc=\{-n_F \Delta f,-(n_F-1) \Delta f, \ldots,0,\ldots,(n_F-1) \Delta f,n_F \Delta f\}$.  Equivalently, $ \Smc = \{f_n\}_{n=1}^{N_F} $ with $ f_n = (n- n_F-1) \Delta f $.
We will approximate (\ref{eq:bke_1}) at the grid points $f_n$ by first approximating the derivative by a centered finite difference at the points $f=f_n+\Delta f/2$ and $f=f_n-\Delta f/2$, 
 \begin{displaymath}
 \frac{\de u_n(t)}{\de t} \approx \ds \frac{\sigma_F^2}{\tau}\e{\beta V(f_n)} \frac{\left[ \e{-\beta V(f)} \partial_f u(f,t) \right]\big|_{f=f_n+\Delta f/2}-\left[ \e{-\beta V(f)} \partial_f u(f,t) \right]\big|_{f=f_n-\Delta f/2} }{\Delta f}, 
\end{displaymath}
where,
\begin{displaymath}
 u_n(t)=u(f_n,t)
\end{displaymath}
is just the point evaluation of the function $u(f,t)$ at the grid point $f=f_n$. Next, we approximate the derivative at $f=f_n\pm \Delta f/2$ once again by a centered difference this time around the grid points $f=f_n\pm \Delta f $ 
and $f=f_n$. The final approximation can be written as,
 \begin{displaymath}
 \frac{\de u_n(t)}{\de t} \approx \ds K_{n,n+1} u_{n+1}(t)-\left(K_{n,n+1}+K_{n,n-1}\right)u_n(t) + K_{n,n-1} u_{n-1}(t),
\end{displaymath}
with,
\begin{eqnarray} \label{eq:transRatesF}
 K_{n,n+1}&=&  \frac{\sigma_F^2}{\tau \Delta f^2} \e{- \beta \left( V(f_n+\Delta f/2)-V(f_n)\right)}, \\
 K_{n,n-1}&=&  \frac{\sigma_F^2}{\tau \Delta f^2} \e{- \beta \left( V (f_n-\Delta f/2)- V (f_n)\right)}. \nonumber
\end{eqnarray}
 The approximation of the backward-Kolmogorov equation can be expressed as,
\begin{displaymath}
\frac{\de}{\dt} \bs{u}(t)=\varmatrix{L} \bs{u}(t),
\end{displaymath}
where the entries of the matrix $\varmatrix{L}$ are given as 
\begin{displaymath}
 [\varmatrix{L}]_{n,\np}=\left\{ \begin{array}{cc}
                          K_{n,n+1} & \textrm{if } \np=n+1, 1 \leq n,\np \leq N_F -1\\
                          K_{n,n-1} & \textrm{if } \np=n-1, 2 \leq n, \np \leq N_F \\
                          - K_{n,n+1} - K_{n,n-1} & \textrm{if } \np=n, 2 \leq n \leq N_F -1\\
                          - K_{1,2} & \textrm{if } \np=n=1\\
                          -K_{N_F,N_F-1} & \textrm{if } \np=n=N_F \\
			  0 & \textrm{otherwise.}
                         \end{array} \right.
\end{displaymath}
which, for any choice of $ \Delta f > 0 $, defines a Markov jump process  with space state defined by the grid points $\{f_n\}_{n=1}^{N_F}$ and jump rates between $f_n$ and $f_{n\pm 1}$ given by $K_{n,n\pm1}$. Moreover, it can be easily checked that the vector,
\begin{displaymath}
 [\boldsymbol{\pi}]_n = \e{-\beta V(f_n)},
\end{displaymath}
solves the equation,
\begin{displaymath}
 \varmatrix{L}^* \boldsymbol{\pi} = 0,
\end{displaymath}
where $\varmatrix{L}^*$ is the adjoint matrix of $\varmatrix{L}$. This means that $\boldsymbol{\pi}$ is an invariant distribution of the Markov jump process and is consistent with the invariant distribution of the continuous process $F(t)$. Moreover, it can also be shown that the Markov jump process satisfies the detailed balance condition with respect to this invariant measure, also consistently with the continuous process. \\
We must next  truncate the infinite state space, and impose boundary conditions. Since the OU-process has a stationary Gaussian distribution $\rho(f) \sim \e{-\beta f^2/2}$, we choose the last grid point $f_{n_F}$ to be such that,
\begin{displaymath}
 \e{-\beta f_{n_F}^2/2}=\delta,
\end{displaymath}
where $\delta$ is a small number. In practice we choose $\delta=10^{-14}$. The discrete-state, Markov jump approximation of the process $\Fmc (t)$ has state space $\Smc=\{-n_F \Delta f,-(n_F-1) \Delta f, \ldots,0,\ldots,(n_F-1) \Delta f,n_F \Delta f\}$ and transition rates given by (\ref{eq:transRatesF}).

\bibliographystyle{model1-num-names}
\bibliography{references}
\end{document}